\documentstyle[epsf]{mystyle}

\makeatletter
\let\chapter\hid@chapter
\makeatother

\def\spose#1{\hbox to 0pt{#1\hss}}
\def\lsim{\mathrel{\spose{\lower 3pt\hbox{$\mathchar"218$}}
 \raise 2.0pt\hbox{$\mathchar"13C$}}}
\def\gsim{\mathrel{\spose{\lower 3pt\hbox{$\mathchar"218$}}
 \raise 2.0pt\hbox{$\mathchar"13E$}}}
\begin{document}

%%%%%%%%%   Make CERN Titlepage   %%%%%%%%%%%%%%%%%
\thispagestyle{empty}

\begin{titlepage}

\begin{flushright}
CERN-TH/98-2\\
hep-ph/9801269
\end{flushright}

\vspace{1.5cm}
\begin{center}
\LARGE\bf
Heavy-Quark Effective Theory and\\
Weak Matrix Elements
\end{center}

\vspace{0.5cm}
\begin{center}
Matthias Neubert\\[0.1cm]
{\sl Theory Division, CERN, CH-1211 Geneva 23, Switzerland}
\end{center}

\vspace{1cm}
\begin{center}
{\bf Abstract:}\\
\vspace{0.2cm}\noindent
\parbox{12cm}{
Recent developments in the theory of weak decays of heavy flavours are
reviewed. Applications to exclusive semileptonic $B$ decays, the
semileptonic branching ratio and charm counting, beauty lifetimes, and
hadronic $B$ decays are discussed.}
\end{center}

\vspace{1.5cm}
\begin{center}
{\sl
Invited talk presented at the\\
International Europhysics Conference on High Energy Physics\\
Jerusalem, Israel, 19--26 August 1997}
\end{center}

\vspace{2cm}
\noindent
CERN-TH/98-2\\
January 1998

\end{titlepage}

\thispagestyle{empty}
\vbox{}
\newpage
\setcounter{page}{1}
%%%%%%%%%   End CERN Titlepage   %%%%%%%%%%%%%%%%%

%\pagenumbering{empty} %%% omit to avoid Lellouch's bug!

\authorrunning{M.\,Neubert}
\titlerunning{{\talknumber}: Heavy-Quark Effective Theory}

% For plenary talks, the talk number is that of the session
\def\talknumber{19}

\title{{\talknumber}:
Heavy-Quark Effective Theory and Weak Matrix Elements}

\author{Matthias Neubert (Matthias.Neubert@cern.ch)}

\institute{Theory Division, CERN, CH-1211 Geneva 23, Switzerland}

\maketitle

\begin{abstract}
Recent developments in the theory of weak decays of heavy flavours are
reviewed. Applications to exclusive semileptonic $B$ decays, the
semileptonic branching ratio and charm counting, beauty lifetimes, and
hadronic $B$ decays are discussed.
\end{abstract}

\section{Introduction and theoretical concepts}

To discuss even the most significant recent theoretical developments in
heavy-flavour physics in a single talk is a difficult task.
Fortunately, this field is blooming and will continue to be of great
importance in view of several new experimental facilities ($B$
factories) to start operating in the near future. Below, I will review
the latest theoretical developments in this field and discuss the most
important phenomenological applications. They concern semileptonic $B$
decays and the measurements of the CKM parameters $|V_{cb}|$ and
$|V_{ub}|$, the semileptonic branching ratio and charm yield in
inclusive $B$ decays, the lifetimes of beauty mesons and baryons, and
hadronic $B$ decays, including the rare decays into two light mesons. I
will start with an introduction to the main theoretical concepts used
in the analysis of these processes.

\begin{figure}
\vspace{-0.5cm}
\begin{center}
\begin{minipage}[t]{4.5cm}
\begin{center}
\epsfxsize=2.5cm\epsffile{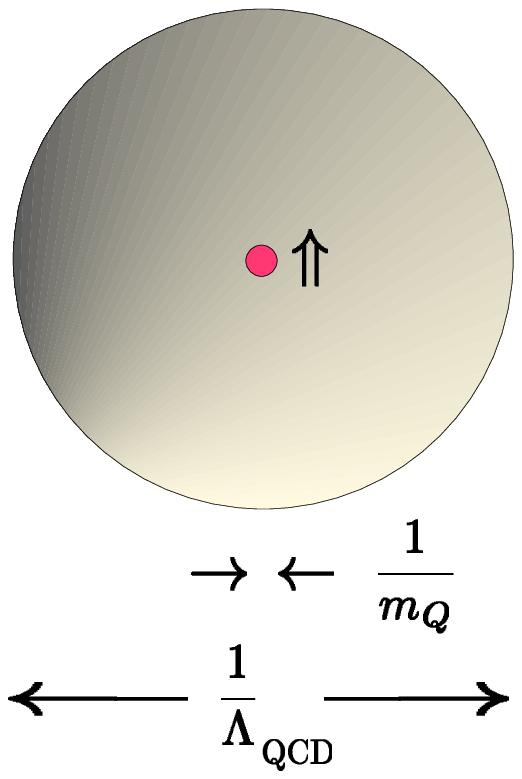}
\end{center}
\end{minipage}
\begin{minipage}[t]{7cm}
\begin{center}
\epsfxsize=7cm\epsffile{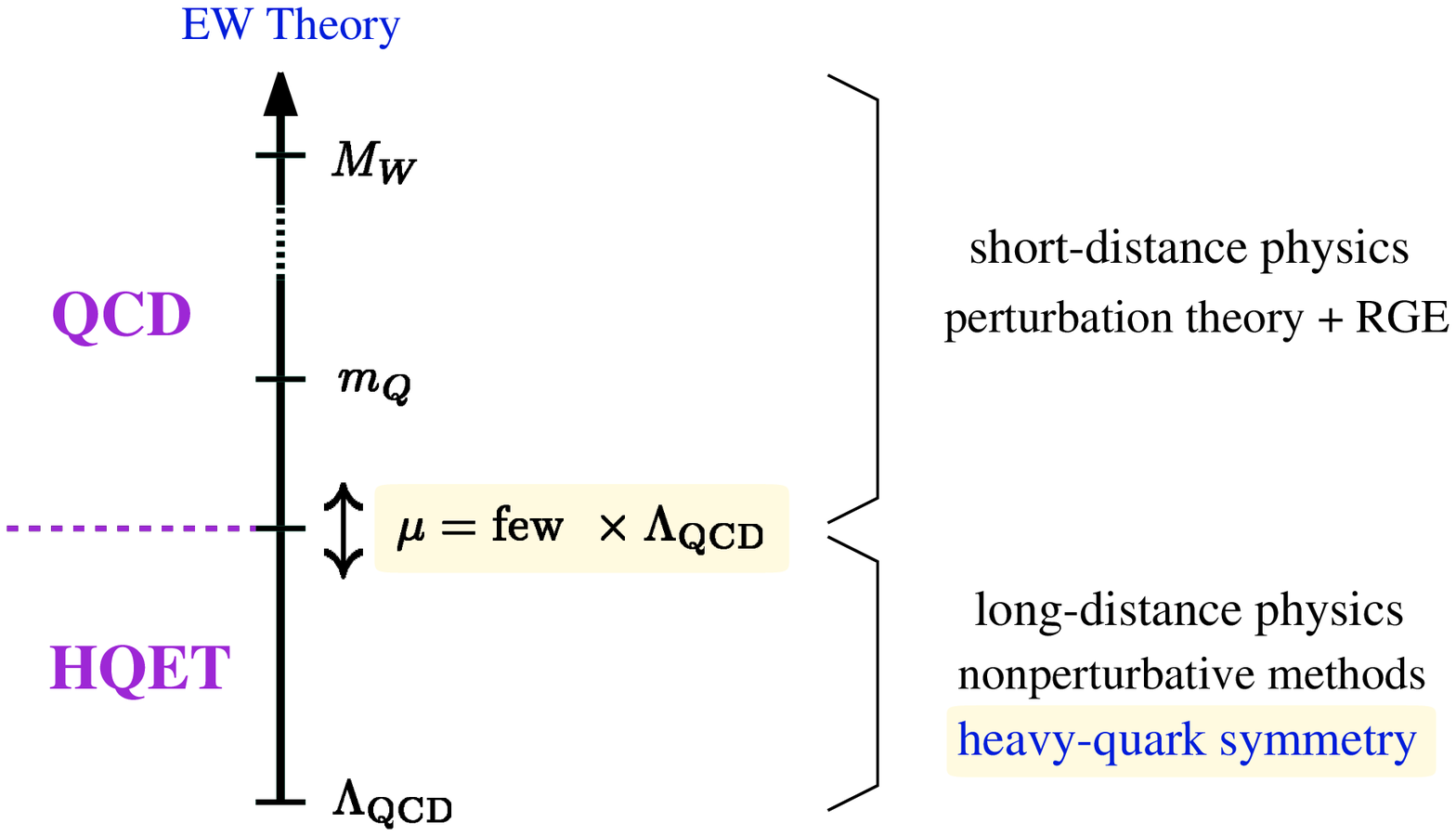}
\end{center}
\end{minipage}
\end{center}
\vspace{-0.5cm}
\caption{Length scales of a heavy hadron, and construction of the HQET}
\label{fig:1and2}
\end{figure}

The properties of hadrons containing a single heavy quark $Q$ are
characterized by the large separation of two length scales: the Compton
wave length $1/m_Q$ of the heavy quark is much smaller than the typical
size $1/\Lambda_{\rm QCD}$ of hadronic bound states in QCD (see the
left plot in Fig.~\ref{fig:1and2}). In the limit $m_Q\to\infty$, the
configuration of the light degrees of freedom in the hadron becomes
independent of the spin and flavour of the heavy quark. In that limit
there is a global SU$(2 n_h)$ spin--flavour symmetry of the strong
interactions, where $n_h$ is the number of heavy-quark flavours
\cite{review}. This symmetry helps in understanding the spectroscopy
and decays of heavy hadrons from first principles. It does not allow us
to solve QCD, but to parametrize the strong-interaction effects of
heavy-quark systems by a minimal number of reduced matrix elements,
thus giving rise to nontrivial relations between observables. In
particular, all form factors for the weak $\bar B\to D^{(*)}\ell\,
\bar\nu$ transitions are proportional to a universal function $\xi(w)$,
where $w=v_B\cdot v_{D^{(*)}}$ is the product of the meson velocities.
At the zero-recoil point $w=1$, corresponding to $v_B=v_{D^{(*)}}$,
this function is normalized to unity: $\xi(1)=1$. The symmetry-breaking
corrections to the heavy-quark limit can be organized in an expansion
in powers of the small parameters $\alpha_s(m_Q)$ and $\Lambda_{\rm
QCD}/m_Q$. A convenient way to do this is provided by the heavy-quark
effective theory (HQET), whose purpose is to separate the short- and
long-distance physics associated with the two length scales, making all
dependence on the large mass scale $m_Q$ explicit. This allows us to
derive scaling laws relating different observables to each other. The
philosophy behind the HQET is illustrated in the right plot in
Fig.~\ref{fig:1and2}.

The effective Lagrangian of the HQET is
\[
   {\cal L}_{\rm eff} = \bar h\,iv\!\cdot\!D h + \frac{1}{2 m_Q}\,
    \bar h\,(iD)^2 h + \frac{C(m_Q)}{4 m_Q}\,\bar h\,\sigma_{\mu\nu}
    G^{\mu\nu} h + \dots \,,
\]
where $v$ is the velocity of the hadron containing the heavy quark, and
$h$ is the heavy-quark field. The Lagrangian incorporates the
spin--flavour symmetry to leading order in $1/m_Q$. The only nontrivial
short-distance coefficient has recently been calculated to
next-to-leading and even next-to-next-to-leading order
\cite{ABN97}--\cite{GrNe97}. The result is
\[
   C(m_Q) = \big[\alpha_s(m_Q)\big]^{9/25} \Big\{ 1
   + 0.672\alpha_s(m_Q) + (1.33\pm 0.04)\alpha_s^2(m_Q)
   + \dots \Big\} \,.
\]
The long-distance physics is encoded in the hadronic matrix elements of
HQET operators, e.g.
\begin{eqnarray}
   \mu_\pi^2(H) &=& - \langle H|\bar h\,(iD)^2 h|H\rangle \,, \qquad
    \mu_G^2(H) = \frac{C(m_Q)}{2}\,\langle H|\bar h\,\sigma_{\mu\nu}
    G^{\mu\nu} h|H\rangle \,, \nonumber\\
   \bar\Lambda(H) &=& M_H - m_Q
    + \frac{\mu_G^2(H) - \mu_\pi^2(H)}{2 M_H} + \dots \,, \nonumber
\end{eqnarray}
which are referred to as the ``kinetic energy'', the ``chromomagnetic
interaction'', and the ``binding energy'', respectively. They are
important parameters entering many applications of the heavy-quark
expansion, the calculation of inclusive decay rates and spectra in
particular. Certain combinations of these parameters can be extracted
from spectroscopy:
\begin{eqnarray}
   \mu_G^2(B) &=& \frac 34\,(M_{B^*}^2-M_B^2) \approx 0.36\,
    {\rm GeV}^2 \,, \qquad \mu_G^2(\Lambda_b) = 0 \,, \nonumber\\
   \mu_\pi^2(\Lambda_b) - \mu_\pi^2(B) &=& -\frac{M_B M_D}{2}
    \left( \frac{M_{\Lambda_b}-M_{\Lambda_c}}{M_B-M_D}
    - \frac 34\,\frac{M_{B^*}-M_{D^*}}{M_B-M_D} - \frac 14 \right)
    \approx 0 \,, \nonumber\\
   \bar\Lambda(\Lambda_b) - \bar\Lambda(B) &\approx& 0.31\,{\rm GeV}
    \,. \nonumber
\end{eqnarray}
The individual parameters $\mu_\pi^2$ and $\bar\Lambda$ are scheme
dependent, however, because of renormalon ambiguities in their
definition. Once a scheme is chosen, they can be calculated
nonperturbatively, or extracted from moments of inclusive decay
spectra. Table~\ref{tab:2} shows a collection of some experimental
determinations of these parameters in the on-shell scheme.

\begin{table}
\vspace{-0.5cm}
\caption{Determinations of the parameters $\bar\Lambda$ and $\mu_\pi^2$
from inclusive decay spectra}
\begin{center}
\begin{tabular}{|l|l|cc|}\hline
{}~Reference & ~Method & ~$\bar\Lambda(B)$ [GeV]~ &
 ~$\mu_\pi^2(B)$ [GeV$^2$]~ \\
\hline
{}~Falk et al.\ \protect\cite{FLS2} & ~Hadron Spectrum~ &
 $\approx 0.45$ & $\approx 0.1$ \\
\hline
{}~Gremm et al.\ \protect\cite{GKLW} & ~Lepton Spectrum &
 $0.39\pm 0.11$ & $0.19\pm 0.10$ \\
{}~Chernyak \protect\cite{Cher} & ~$(\bar B\to X\,\ell\,\bar\nu)$ &
 $0.28\pm 0.04$ & $0.14\pm 0.03$ \\
{}~Gremm, Stewart \protect\cite{GrSt96}~ & & $0.33\pm 0.11$ &
 $0.17\pm 0.10$ \\
\hline
{}~Li, Yu \protect\cite{LiYu} & ~Photon Spectrum &
 $0.65_{-0.30}^{+0.42}$ & $0.71_{-0.70}^{+1.16}$ \\
 & ~$(\bar B\to X_s\gamma)$ & & \\
\hline
\end{tabular}
\end{center}
\vspace{-0.3cm}
\label{tab:2}
\end{table}

Using the operator product expansion, any inclusive decay rate of a
beauty hadron can be expanded as
\[
   \Gamma(H) = \frac{G_F^2 m_b^5}{192\pi^3}
   \left( 1 - \frac{\mu_\pi^2(H)}{2 m_b^2} \right)
   \left\{ c_3 + c_5\,\frac{\mu_G^2(H)}{m_b^2} + \sum_n\,
   c_6^{(n)}\,\frac{\langle O_n\rangle_H}{m_b^3} + \dots \right\} \,,
\]
where $\langle O_n\rangle_H$ are the matrix elements of local
four-quark operators, which parametrize nonspectator effects in these
decays, and $c_i$ are calculable short-distance coefficients, which
depend on CKM parameters, the ratios of quark masses, and the
renormalization scheme. The free quark decay emerges as the leading
term in a systematic $1/m_b$ expansion, with bound-state corrections
suppressed by two inverse powers of the heavy-quark mass. Note that
ratios of inclusive decay rates are independent of $\mu_\pi^2$ and the
common factor $m_b^5$, as well as of most CKM parameters. The
application of the operator product expansion to the calculation of
inclusive decay rates relies on the assumption of quark--hadron
duality. Strictly speaking, the theoretical description of such
processes is thus not entirely from first principles.

\section{Exclusive semileptonic decays}

The most important applications of the HQET concern the description of
exclusive semileptonic decays based on the quark transition $b\to
c\,\ell\,\bar\nu$. This is where the theory is well tested, and
theoretical uncertainties are best understood. The most important
result is a precision determination of the CKM parameter $|V_{cb}|$
\cite{Vcb}.

\subsection{Determination of $|V_{cb}|$ from $\bar{B}\to
D^{(*)}\ell\,\bar{\nu}$ decays}

The differential semileptonic decay rates as a function of the
kinematical variable $w=v_B\cdot v_{D^{(*)}}$ are given by
\begin{eqnarray}
   \frac{{\rm d}\Gamma(\bar B\to D^*\ell\,\bar\nu)}{{\rm d}w}
   &=& \frac{G_F^2 M_B^5}{48\pi^3}\,r_*^3 (1-r_*)^2\,\sqrt{w^2-1}\,
    (w+1)^2 \nonumber\\
   &&\times \left[ 1 + \frac{4w}{w+1}\,\frac{1-2w r_*+r_*^2}{(1-r_*)^2}
    \right]\,|V_{cb}|^2\,{\cal F}^2(w) \,, \nonumber\\
   \frac{{\rm d}\Gamma(\bar B\to D\,\ell\,\bar\nu)}{{\rm d}w}
   &=& \frac{G_F^2 M_B^5}{48\pi^3}\,r^3 (1+r)^2\,(w^2-1)^{3/2}\,
    |V_{cb}|^2\,{\cal G}^2(w) \,, \nonumber
\end{eqnarray}
where $r_{(*)}=M_{D^{(*)}}/M_B$. In the heavy-quark limit, the form
factors ${\cal F}(w)$ and ${\cal G}(w)$ coincide with the universal
function $\xi(w)$ and are thus normalized to unity at $w=1$. Much
effort has gone into calculating the symmetry-breaking corrections to
this limit, with the result that \cite{BaBar}
\begin{eqnarray}
   {\cal F}(1) &=& 1 + c_A(\alpha_s) + 0 + \delta_{1/m^2} + \dots
    = 0.924\pm 0.027 \,, \nonumber\\
   {\cal G}(1) &=& 1 + c_V(\alpha_s) + \delta'_{1/m}
    + \delta'_{1/m^2} + \dots = 1.00\pm 0.07 \,, \nonumber
\end{eqnarray}
where the short-distance coefficients $c_A$ and $c_V$ are known to
two-loop order \cite{Cz97,Franz}. These numbers include the
leading-logarithmic QED corrections. The absence of first-order power
corrections to ${\cal F}(1)$ is a consequence of Luke's theorem
\cite{Luke}. The theoretical errors quoted above include the
perturbative uncertainty and the uncertainty in the calculation (and
truncation) of power corrections, added in quadrature. If instead the
errors are added linearly, the result for ${\cal F}(1)$ changes to
$0.924\pm 0.041$. This value has recently been confirmed in a different
regularization scheme, in which the separation between short- and
long-distance contributions is achieved by means of a hard momentum
cutoff \cite{Ur97}.

A value for $|V_{cb}|$ can be obtained by extrapolating experimental
data for the differential decay rates to the zero-recoil point, using
theoretical constraints on the shape of the form factors.
Model-independent bounds on the physical $\bar B\to
D^{(*)}\ell\,\bar\nu$ form factors can be derived using analyticity
properties of QCD correlators, unitarity and dispersion relations
\cite{BoydVcb,CN96}. Combining these methods with the approximate
heavy-quark symmetry, very powerful one-parameter functions can be
derived, which approximate the physical form factors in the
semileptonic region with an accuracy of better than 2\% \cite{CLN97}.
For instance, the function ${\cal G}(w)$ can be parametrized as
\[
   \frac{{\cal G}(w)}{{\cal G}(1)} \approx 1 - 8 \rho_1^2 z(w)
    + (51.\rho_1^2 - 10.) z^2(w) - (252.\rho_1^2-84.) z^3(w) \,,
\]
where $z(w)=(\sqrt{w+1}-\sqrt 2)/(\sqrt{w+1}+\sqrt 2)$, and $\rho_1^2$
is the (negative) slope of the form factor at zero recoil. A similar
parametrization can be given for the function ${\cal F}(w)$. At
present, these constraints are only partially included in the analyses
of experimental data.

The world-average results of such analyses are \cite{Persis,GibbWars}
\begin{eqnarray}
   |V_{cb}|\,{\cal F}(1) &=& (35.2\pm 2.6)\times 10^{-3} \,,
    \nonumber\\
   |V_{cb}|\,{\cal G}(1) &=& (38.6\pm 4.1)\times 10^{-3} \,. \nonumber
\end{eqnarray}
A good fraction of the present errors reflect the uncertainty in the
extrapolation to zero recoil, which could be avoided by implementing
the dispersive constraints mentioned above. When combined with the
theoretical predictions for the normalization of the form factors at
zero recoil, the data yield the accurate value
\[
   |V_{cb}| = (38.2\pm 2.3_{\rm exp}\pm 1.2_{\rm th})\times 10^{-3} \,,
\]
which is in good agreement with an independent determination from
inclusive $B$ decays \cite{review,Persis}.

\subsection{Tests of heavy-quark symmetry}

\begin{figure}
\vspace{-0.6cm}
\epsfxsize=9cm
\centerline{\epsffile{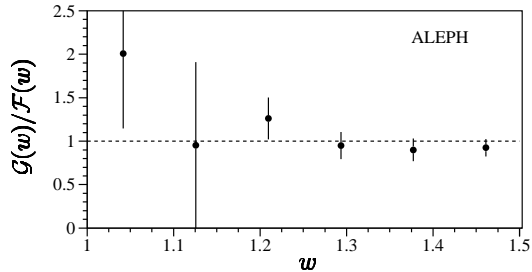}}
\vspace{-4.7cm}
\caption{Ratio of the two form factors ${\cal G}(w)$ and ${\cal
F}(w)$, from Ref.~\protect\cite{ALEPHGF}}
\label{fig:4}
\end{figure}

In general, the decays $\bar B\to D\,\ell\,\bar\nu$ and $\bar B\to
D^*\ell\,\bar\nu$ are described by four independent form factors:
${\cal G}(w)$ for the former process, and $h_{A1}(w)$, $R_1(w)$,
$R_2(w)$ -- a combination of which defines the function ${\cal F}(w)$
-- for the latter one. In the heavy-quark limit, ${\cal G}(w)$, ${\cal
F}(w)$ and $h_{A1}(w)$ become equal to the function $\xi(w)$, whereas
$R_1(w)$ and $R_2(w)$ approach unity. The universality of the function
$\xi(w)$ can be tested by measuring the ratio ${\cal G}(w)/{\cal F}(w)$
as a function of $w$. The ALEPH data for this ratio \cite{ALEPHGF} are
shown in Fig.~\ref{fig:4}; a similar measurement has also been reported
by CLEO \cite{CLEOGF}. Within errors, the data are compatible with a
universal form factor. At large recoil, where the experimental errors
are smallest, this provides a test of heavy-quark symmetry at the level
of 10--15\%.

A more refined analysis of symmetry-breaking corrections has been done
by measuring the ratios $R_1$ and $R_2$ close to zero recoil. There the
HQET predicts that \cite{review}
\begin{eqnarray}
   R_1 &\approx& 1 + \frac{4\alpha_s(m_c)}{3\pi}
    + \frac{\bar\Lambda}{2 m_c} = 1.3\pm 0.1 \,, \nonumber\\
   R_2 &\approx& 1 - \frac{\bar\Lambda}{2 m_c} = 0.8\pm 0.2 \,,
    \nonumber\\
   \rho_{A_1}^2 - \rho_{\cal F}^2 &=& \frac{R_1-1}{6}
    + \frac{1-R_2}{3(1-r_*)} = 0.2\pm 0.1 \,. \nonumber
\end{eqnarray}
The CLEO data for these three quantities are $R_1=1.24\pm 0.29$,
$R_2=0.72\pm 0.19$, and $\rho_{A_1}^2-\rho_{\cal F}^2=0.20\pm 0.19$
\cite{CLEOtests}, in good agreement with the theoretical predictions.
With a little more precision, the data would start to test the pattern
of symmetry-breaking effects in the heavy-quark expansion.

\subsection{Total exclusive semileptonic rates}

Summing up the average experimental results $\mbox{B}(\bar B\to
D\,\ell\,\bar\nu)=(1.95\pm 0.27)\%$, $\mbox{B}(\bar B\to
D^*\ell\,\bar\nu)=(5.05\pm 0.25)\%$, $\mbox{B}(\bar B\to
D^{(*)}\pi\,\ell\,\bar\nu)=(2.30\pm 0.44)\%$, and adding $\mbox{B}(\bar
B\to X_u\,\ell\,\bar\nu)=(0.15\pm 0.10)\%$ for the inclusive branching
ratio for semileptonic decays into charmless final states, one gets a
total of $(9.45\pm 0.58)\%$ \cite{Persis}, which is not far below the
average value for the inclusive semileptonic branching ratio,
$\mbox{B}(\bar B\to X\,\ell\,\bar\nu)=(10.19\pm 0.37)\%$, measured at
the $\Upsilon(4s)$ resonance (see below). Thus, there is little room
for extra contributions.

A solid theoretical understanding of $B$ decays into p-wave charm meson
resonances would be important in order to address the question of
whether there are additional contributions not accounted for in the
above sum, and also to understand the main source of background in the
determination of $|V_{cb}|$ from $\bar B\to D^*\ell\,\bar\nu$ decays.
The description of these processes in the context of the HQET involves,
at leading order, two new universal functions: $\tau_{3/2}(w)$ for the
decays into the narrow states $(D_1,D_2^*)$, and $\tau_{1/2}(w)$ for
those into the broad states $(D_0^*,D_1^*)$. In the heavy-quark limit,
the differential (in $w$) decay rates vanish at zero recoil, since in
that limit the p-wave states are orthogonal to the ground state. There
is a sizable $1/m_c$ correction to the $\bar B\to D_1\,\ell\,\bar\nu$
decay rate at zero recoil, which can be calculated in a
model-independent way in terms of known charm meson masses \cite{Leib}.
It is important, since the kinematical region is restricted close to
zero recoil ($1<w<1.3$). Detailed theoretical predictions for the
semileptonic and hadronic decay rates into p-wave charm states, which
incorporate the constraints imposed by heavy-quark symmetry, can be
found in Refs.~\cite{Leib,MNpwave}.

\boldmath
\section{Semileptonic $b\to u$ decays and $|V_{ub}|$}
\unboldmath

Two exclusive semileptonic $B$ decay modes into charmless hadrons have
been observed by CLEO; the corresponding branching ratios are
\cite{CLEOVub}
\begin{eqnarray}
   \mbox{B}(\bar B\to\pi\,\ell\,\bar\nu) &=& (1.8\pm 0.5_{\rm exp}
    \pm 0.2_{\rm model})\times 10^{-4} \,, \nonumber\\
   \mbox{B}(\bar B\to\rho\,\ell\,\bar\nu)
   &=& (2.5_{-0.8{\rm exp}}^{+0.6}\,\pm 0.5_{\rm model})
    \times 10^{-4} \,. \nonumber
\end{eqnarray}
That the theoretical description of these processes involves
heavy-to-light form factors implies a certain amount of model
dependence, since heavy-quark symmetry does not help to fix their
normalization in a precise way. To some extent, a discrimination
between models can be obtained by requiring a simultaneous fit of both
exclusive channels. {}From a $\chi^2$-weighted average of models, CLEO
obtains \cite{CLEOVub}
\[
   |V_{ub}| = (3.3\pm 0.4_{\rm exp}\pm 0.7_{\rm model})
   \times 10^{-3} \,.
\]
Even with the present, very limited statistics of the measurements, the
theoretical uncertainties are the limiting factor in the determination
of $|V_{ub}|$. In the future, the model dependence can and will be
reduced mainly by combining different theoretical approaches, in
particular: lattice calculations, which are restricted to the region of
large $q^2$ \cite{lattVub}; light-cone QCD sum rules, including
$O(\alpha_s)$ and higher-twist corrections \cite{SRVub,BB97};
analyticity and unitarity constraints \cite{dispVub,BoydVub};
dispersion relations \cite{BurdVub}. Although the optimal strategy is
not yet clear at present, I believe it will be possible to reach the
level of 15\% theoretical uncertainty.

\begin{figure}
\vspace{-0.5cm}
\begin{center}
\begin{minipage}[t]{5.8cm}
\begin{center}
\epsfxsize=5.8cm\epsffile{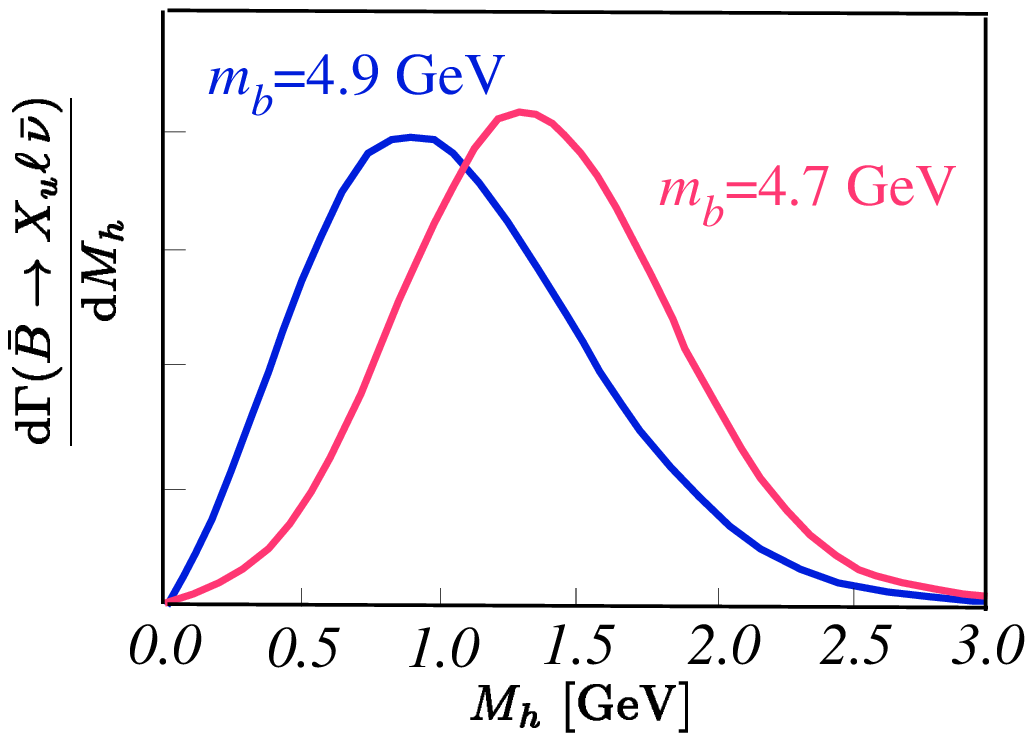}
\end{center}
\end{minipage}
\begin{minipage}[t]{5.8cm}
\begin{center}
\epsfxsize=5.8cm\epsffile{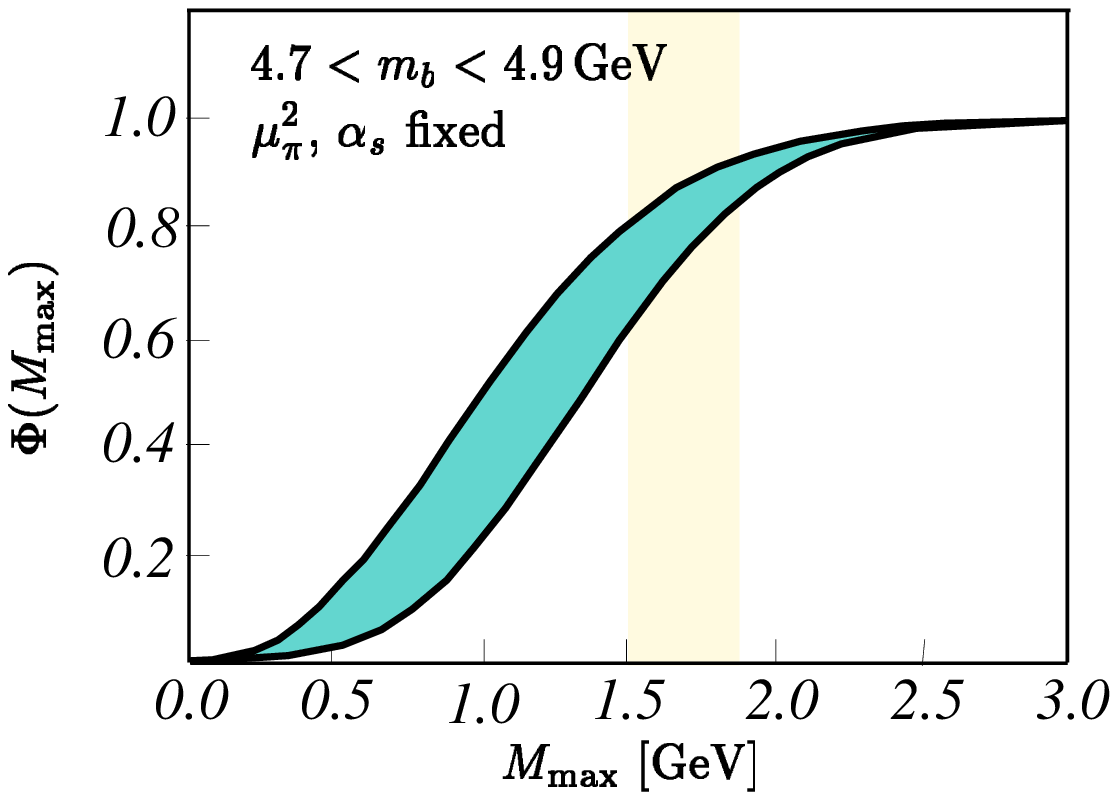}
\end{center}
\end{minipage}
\end{center}
\vspace{-0.5cm}
\caption{Hadronic mass distribution (left) and fraction of $\bar B\to
X_u\,\ell\,\bar\nu$ decays with invariant hadronic mass below $M_{\rm
max}$ (right), from Ref.~\protect\cite{Dike97}}
\label{fig:6}
\end{figure}

The traditional way to extract $|V_{ub}|$ from inclusive $\bar B\to
X\,\ell\,\bar\nu$ decays has been to look at the endpoint region of the
charged-lepton energy spectrum, where there is a tiny window not
accessible to $\bar B\to X_c\,\ell\,\bar\nu$ decays. However, this
method involves a large extrapolation and is plagued by uncontrolled
(and often underestimated) theoretical uncertainties. A reanalysis of
the available experimental data (using the ISGW2 model) gives $|V_{ub}|
= (3.7\pm 0.6_{\rm exp})\times 10^{-3}$ \cite{Gibbons}, in agreement
with the value quoted above. A better discrimination between $b\to u$
and $b\to c$ transitions should use vertex information combined with a
cut on the invariant mass $M_h$ (or energy $E_h$) of the hadronic final
state \cite{Barg90}--\cite{CJin}. Parton model calculations (with Fermi
motion included) indicate that about 90\% of all $\bar B\to
X_u\,\ell\,\bar\nu$ decays have $M_h<M_D$, as shown in
Fig.~\ref{fig:6}. Ideally, this cut would thus provide for a very
efficient discriminator. In practise, there will be some leakage so
that presumably one will be forced to require $M_h<M_{\rm max}$ with
some threshold $M_{\rm max}<M_D$. The task for theorists is to
calculate the fraction of events with hadronic mass $M_h<M_{\rm max}$,
\[
   \Phi(M_{\rm max}) = \frac{1}{\Gamma} \int\limits_0^{M_{\rm max}}\!
   \mbox{d}M_h\,\frac{{\rm d}\Gamma(\bar B\to X_u\,\ell\,\bar\nu)}
                     {{\rm d}M_h} \,.
\]
To calculate this fraction requires an ansatz for the ``shape
function'', which describes the Fermi motion of the $b$ quark inside
the $B$ meson \cite{MN94,BSUV94}. The first three moments of this
function are determined in terms of known HQET matrix elements. Still,
some theoretical uncertainty remains, mainly associated with the values
of the $b$-quark mass and the kinetic energy $\mu_\pi^2$, as well as
unknown $O(\alpha_s^2)$ corrections. As an example, Fig.~\ref{fig:6}
shows the dependence of $\Phi(M_{\rm max})$ on the value of $m_b$. It
turns out that the resulting theoretical uncertainty strongly depends
on the value of the threshold $M_{\rm max}$. First estimates yield
$\delta|V_{ub}|/|V_{ub}|\approx 10\%$ for $M_{\rm max}=M_D$, and
$\delta|V_{ub}|/|V_{ub}|\approx 20\%$ for $M_{\rm max}=1.5\,$GeV
\cite{Dike97,FLW97}. This new method is challenging both for theorists
and for experimenters, but it is superior to the endpoint method. I
believe that ultimately a theoretical accuracy of 10\% can be reached.

\section{Semileptonic branching ratio and charm multiplicity}

\begin{figure}
\vspace{-0.5cm}
\epsfxsize=6.5cm
\centerline{\epsffile{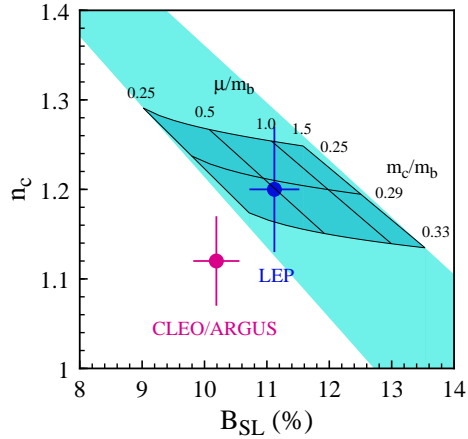}}
\vspace{-0.2cm}
\caption{Theory \protect\cite{NS97} versus experiment for the
semileptonic branching ratio and charm multiplicity}
\label{fig:7}
\end{figure}

For many years, the apparent discrepancy between (some) measurements of
the semileptonic branching ratio of $B$ mesons and theoretical
predictions for this quantity has given rise to controverse dispute and
speculations about deviations from the Standard Model. Another
important input to this discussion has been the charm multiplicity
$n_c$, i.e.\ the average number of charm (or anticharm) quarks in the
hadronic final state of a $B$ decay. From the theoretical point of
view,
\begin{eqnarray}
   {\rm B_{SL}} &=& \frac{\Gamma(\bar B\to X\,e\,\bar\nu)}
    {\sum_\ell\,\Gamma(\bar B\to X\,\ell\,\bar\nu) + \Gamma_{\rm had}
     + \Gamma_{\rm rare}} \,, \nonumber\\
   n_c &=& 1 + {\rm B}(\bar B\to X_{c\bar c})
    - {\rm B}(\bar B\to\mbox{no~charm}) \nonumber
\end{eqnarray}
are governed by the same partial inclusive decay rates. The theoretical
predictions for these two quantities depend mainly on two parameters:
the quark-mass ratio $m_c/m_b$ and the renormalization scale $\mu$
\cite{Baga,NS97}. The latter dependence reflects our ignorance about
higher-order QCD corrections to the decay rates. The results obtained
by allowing reasonable ranges for these parameters are represented by
the dark-shaded area in Fig.~\ref{fig:7}. The two data points show the
average experimental results obtained from experiments operating at the
$\Upsilon(4s)$ resonance: B$_{\rm SL}=(10.19\pm 0.37)\%$ and
$n_c=1.12\pm 0.05$, and at the $Z$ resonance: B$_{\rm SL}=(11.12\pm
0.20)\%$ and $n_c=1.20\pm 0.07$ \cite{Persis,Michael}. At this
conference, it has been emphasized that a dedicated reanalysis of the
LEP data for the semileptonic branching ratio is necessary, because
some sources of systematic errors had previously been underestimated
\cite{Michael}. To account for this, I have doubled the corresponding
error bar on B$_{\rm SL}$ in Fig.~\ref{fig:7}. If we ignore the LEP
point for the moment, it appears that the theoretical predictions for
both ${\rm B_{SL}}$ and $n_c$ lie significantly higher than the
experimental results.

At this conference, several new ingredients have been presented that
shed some light on this problem. First, there are some nontrivial tests
of the theory. Table~\ref{tab:3} shows details of the theoretical
calculation in the form of the dominant partial decay rates normalized
to ${\rm B}(\bar B\to X_c\,\ell\,\bar\nu)$, for $m_c/m_b=0.29\pm 0.03$,
using the results of Refs.~\cite{Baga,NS97,Uli97}. (Note that the
values for $r_{c\bar ud'}$ are larger than the value $4.0\pm 0.4$ often
used in the literature.) It has been argued that the assumption of
local quark--hadron duality, which underlies the theoretical treatment
of inclusive decay rates, may fail for the decays $\bar B\to X_{c\bar
cs'}$, because there is only little kinetic energy released to the
final state. To test this hypothesis, one can eliminate the
corresponding partial decay rate, in which case one obtains a linear
relation between $n_c$ and ${\rm B_{SL}}$ \cite{Buch95}. The result is
shown as the light band in Fig.~\ref{fig:7}. The fact that, within this
band, the original prediction (i.e.\ the dark-shaded area) is closest
to the data indicates that there is no problem with quark--hadron
duality. Another important check is provided by an experimental
determination of the ratio $r_{c\bar ud'}$ using flavour-specific
measurements of charm branching ratios \cite{CLEOwrongc}. The result is
\cite{Du97}--\cite{Tom97}
\[
   r_{c\bar ud'} =
   \frac{{\rm B}(\bar B\to\mbox{open~$c$}) - {\rm B}(\bar B\to
         \mbox{open~$\bar c$})}{\rm B_{SL}}
   - (2 + r_{c\tau\bar\nu} - r_{u\bar cs'}) = 4.1\pm 0.7 \,,
\]
in good agreement with the theoretical predictions given in
Table~\ref{tab:3}. The theory input $r_{c\tau\bar\nu}-r_{u\bar
cs'}=0.19\pm 0.03$ in this extraction has a very small uncertainty. In
summary, it appears that the heavy-quark expansion works well for both
of the hadronic decay rates.

\begin{table}
\vspace{-0.5cm}
\caption{Predictions for ratios of partial inclusive decay rates}
\begin{center}
\begin{tabular}{|l|cccc|}\hline
 & $r_{c\tau\bar\nu}$ & $r_{c\bar ud'}$ & $r_{c\bar cs'}$ &
 $r_{\rm no\,charm}$ \\
\hline
{}~$\mu=m_b$ & ~$0.22\mp 0.03$~ & ~$4.21\pm 0.01$~ &
 ~$1.89\mp 0.44$~ & ~$0.14\pm 0.04$~ \\
{}~$\mu=m_b/2$~ & ~$0.23\mp 0.03$~ & ~$4.75\pm 0.02$~ &
 ~$2.20\mp 0.49$~ & ~$0.19\pm 0.04$~ \\
\hline
\end{tabular}
\end{center}
\vspace{-0.5cm}
\label{tab:3}
\end{table}

\subsection{Is there a ``missing charm puzzle''?}

Several suggestions have been made to explain why the experimental
value of $n_c=1.12\pm 0.05$ measured by CLEO \cite{Persis} is smaller
than the theoretical prediction $n_c=1.20\pm 0.06$ \cite{NS97}. I have
discussed above that the problem cannot be blamed on violations of
quark--hadron duality, as was originally speculated by some authors. An
interesting new proposal made by Dunietz et al.~\cite{Du97} is based on
the fact that the theoretical definition of $n_c$ refers to a fully
inclusive quantity counting the number of charm and anticharm quarks
per $B$ decay, irrespective of whether they end up as ``open'' or
``hidden'' charm. If there were a sizable branching ratio for decays
into hadronic final states containing undetected $(c\bar c)$ pairs,
\[
   {\rm B}(\bar B\to(c\bar c)_{\rm undetected}+X)\equiv b_{(c\bar c)}
\,,
\]
then the experimentally observed value of $n_c$ would be lower than the
theoretical one: $n_c^{\rm obs}=n_c^{\rm th}-2 b_{(c\bar c)}$. Note
that the conventional charmonium states $J/\psi$, $\psi'$, $\chi_{c1}$,
$\chi_{c2}$, $\eta_c$ are included in the charm counting and thus do
not contribute to $b_{(c\bar c)}$. However, it has been speculated that
there may exist an enhanced production of exotic $(c\bar c g)$ hybrids,
which decay into noncharmed light mesons and could yield a value of
$b_{(c\bar c)}$ of about 10\% \cite{Du97,Close97}.

Alternatively, enhanced flavour-changing neutral current (FCNC)
processes such as $b\to s g$ could, simultaneously, lower the
predictions for ${\rm B_{SL}}$ and $n_c$ by a factor of $[1+{\rm
B}(\mbox{new~FCNC})]^{-1}$, thus providing for a ``new physics
explanation'' of the missing charm puzzle \cite{Alex95,RoSh96}.

\begin{figure}
\vspace{-0.5cm}
\epsfxsize=8cm
\centerline{\epsffile{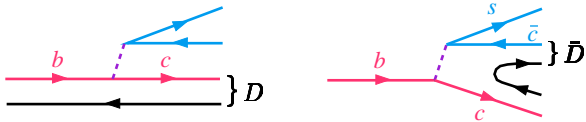}}
\vspace{-0.2cm}
\caption{\label{fig:8}
Mechanisms for $D$ and $\bar D$ production in $\bar B$-meson decays}
\end{figure}

At least a partial answer to the question of whether there is any
support for such speculations is provided by new, flavour-specific
measurements of charm production. Using charm-particle--lepton
correlations in tagged $B$ decays, it is possible to measure the
relative rate of the two mechanisms shown in Fig.~\ref{fig:8}. Whereas
it was previously assumed that the first mechanism was responsible for
all $D$-meson production, recent measurements indicate a rather large
``wrong-charm'' yield \cite{CLEOwrongc}:
\[
   \frac{{\rm B}(\bar B\to\bar D\,X)}{{\rm B}(\bar B\to D\,X)}
   = 0.100\pm 0.026\pm 0.016 \,,
\]
corresponding to a ``wrong-charm'' branching ratio ${\rm B}(\bar
B\to\bar D\,X)=(7.9\pm 2.2)\%$. This observation is supported by
measurements of both exclusive \cite{CLDDbar} and inclusive
\cite{ALDDbar,DELDDbar} production of $D\bar D$ meson pairs. This
effect was not included in Monte Carlo $B$-decay generators and may be
partly responsible for the discrepancy in the CLEO and LEP measurements
of ${\rm B_{SL}}$ and $n_c$ \cite{Du97}. When combined with previously
measured decay rates, the new CLEO result implies \cite{CLEOwrongc}
\[
   {\rm B}(\bar B\to X_{c\bar c}) - b_{(c\bar c)}
   = (21.9\pm 3.7)\% \,,
\]
which is close to the theoretical prediction ${\rm B}(\bar B\to
X_{c\bar c})=(22\pm 6)\%$ \cite{NS97}, indicating that there is not
much room for decays into undetected $(c\bar c)$ pairs, represented by
$b_{(c\bar c)}$. This measurement is supported by a DELPHI result on
the production of two open charm particles \cite{DELbccs}. Furthermore,
by measuring a double ratio of flavour-specific rates, CLEO obtains a
bound on charmless modes that is largely independent of detection
efficiencies and charm branching ratios. The result is ${\rm B}(\bar
B\to\mbox{no~open~charm})=(3.2\pm 4.0)\%<9.6\%$ (90\% CL), which after
subtraction of the known charmonium contributions implies
\cite{CLEOwrongc}
\[
   {\rm B}(\bar B\to\mbox{no~charm}) + b_{(c\bar c)}
   = (0.2\pm 4.1)\% < 6.8\% \quad (90\%~\mbox{CL}) \,.
\]
Given that the Standard Model prediction for the charmless rate is
${\rm B}(\bar B\to\mbox{no~charm})=(1.6\pm 0.8)\%$ \cite{Uli97}, it
appears that there is little room for either hidden $(c\bar c)$
production or new physics contributions. Again, this conclusion is
supported by a DELPHI result obtained using impact parameter
measurements: ${\rm B}(\bar B\to\mbox{no~open~charm})=(4.5\pm
2.5)\%<8.4\%$ (90\% CL) \cite{DELbccs}.

Combining these two measurements, I conclude that $b_{(c\bar c)}<5\%$
(90\% CL) cannot be anomalously large. The same bound applies to other,
nonstandard sources of charmless $B$ decays, i.e.\ ${\rm
B}(\mbox{new~FCNC})<5\%$ (90\% CL). This conclusion is supported by the
DELPHI limit ${\rm B}(\bar B\to X_{sg})<5\%$ (95\% CL) obtained from a
study of the $p_\perp$ spectrum of charged kaons produced in $B$ decays
\cite{DELbccs}. It must be noted, however, that a preliminary
indication of a kaon excess at large $p_\perp$, as expected from
enhanced $b\to s g$ transitions \cite{Ra97}, has been reported by SLD
at this conference \cite{SLDnew}. Even though a definite conclusion can
therefore not been drawn before these measurements become final, at
present there is no compelling experimental evidence of any
nonstandard physics in inclusive $B$ decays.

The two CLEO measurements quoted above can be combined to give a new
determination of the charm multiplicity, in which the unknown quantity
$b_{(c\bar c)}$ cancels out. The result
\[
   n_c = 1 + \Big[ {\rm B}(\bar B\to X_{c\bar c}) - b_{(c\bar c)} \Big]
   - \Big[ {\rm B}(\bar B\to\mbox{no~charm}) + b_{(c\bar c)} \Big]
   = 1.22\pm 0.06
\]
is significantly higher than (though consistent with) the value
$1.12\pm 0.05$ obtained using the conventional method of charm
counting, and in excellent agreement with the theoretical prediction. I
believe that, ultimately, this new way of measuring $n_c$ will be less
affected by systematic uncertainties than the traditional one. It seem
that the ``missing charm puzzle'' is about to disappear.

\section{Beauty lifetime ratios}

The current world-average experimental results for the lifetime ratios
of different beauty hadrons are \cite{lifetimes}
\[
   \frac{\tau(B^-)}{\tau(B^0)} = 1.06\pm 0.04 \,, \quad
   \frac{\tau(B_s)}{\tau(B_d)} = 0.98\pm 0.07 \,, \quad
   \frac{\tau(\Lambda_b)}{\tau(B^0)} = 0.78\pm 0.04 \,.
\]
Theory predicts that $|\tau(B_s)/\tau(B_d)-1|<1\%$ \cite{BBD96}, and it
will be very difficult to push the experimental accuracy to a level
where one would become sensitive to sub-1\% effects. The theoretical
predictions for the other two ratios have been analysed to third order
in the heavy-quark expansion. Unfortunately, these predictions depend
on some yet unknown ``bag parameters'' $B_1$, $B_2$, $\varepsilon_1$,
$\varepsilon_2$, $\widetilde B$, $r$ parametrizing the hadronic matrix
elements of local four-quark operators. In terms of these parameters,
the results are \cite{NS97}
\begin{eqnarray}
   \frac{\tau(B^-)}{\tau(B^0)} &=& 1 + 16\pi^2\,
    \frac{f_B^2 M_B}{m_b^3}\,\Big[ k_1 B_1 + k_2 B_2
    + k_3\varepsilon_1 + k_4\varepsilon_2 \Big] \,, \nonumber\\
   \frac{\tau(\Lambda_b)}{\tau(B^0)} &=& 0.98 + 16\pi^2\,
    \frac{f_B^2 M_B}{m_b^3}\,\Big[ p_1 B_1 + p_2 B_2
    + p_3\varepsilon_1 + p_4\varepsilon_2
    + (p_5 + p_6\widetilde B) r \Big] \,, \nonumber
\end{eqnarray}
where $k_i$ and $p_i$ are short-distance coefficients, whose values
depend on the ratio $m_c/m_b$ and on the renormalization scale. The
large-$N_c$ counting rules of QCD imply that $B_i=O(1)$ and
$\varepsilon_i=O(1/N_c)$. The factorization approximation for the meson
matrix elements suggests that $B_i\approx 1$ and $\varepsilon_i\approx
0$ \cite{Blok94}. Similarly, the constituent quark model suggests that
$\widetilde B\approx 1$ and $r\approx |\psi_{qq}^{\Lambda_b}(0)|^2/
|\psi_q^B(0)|^2$. The parameter $r$ is the most uncertain one entering
the predictions for the lifetime ratios. Existing theoretical estimates
for this parameter range from 0.1 to 2. Some recent estimates can be
found in Refs.~\cite{Rosn96,Cola96}.

\begin{figure}
%\vspace{-0.5cm}
\epsfxsize=10cm
\centerline{\epsffile{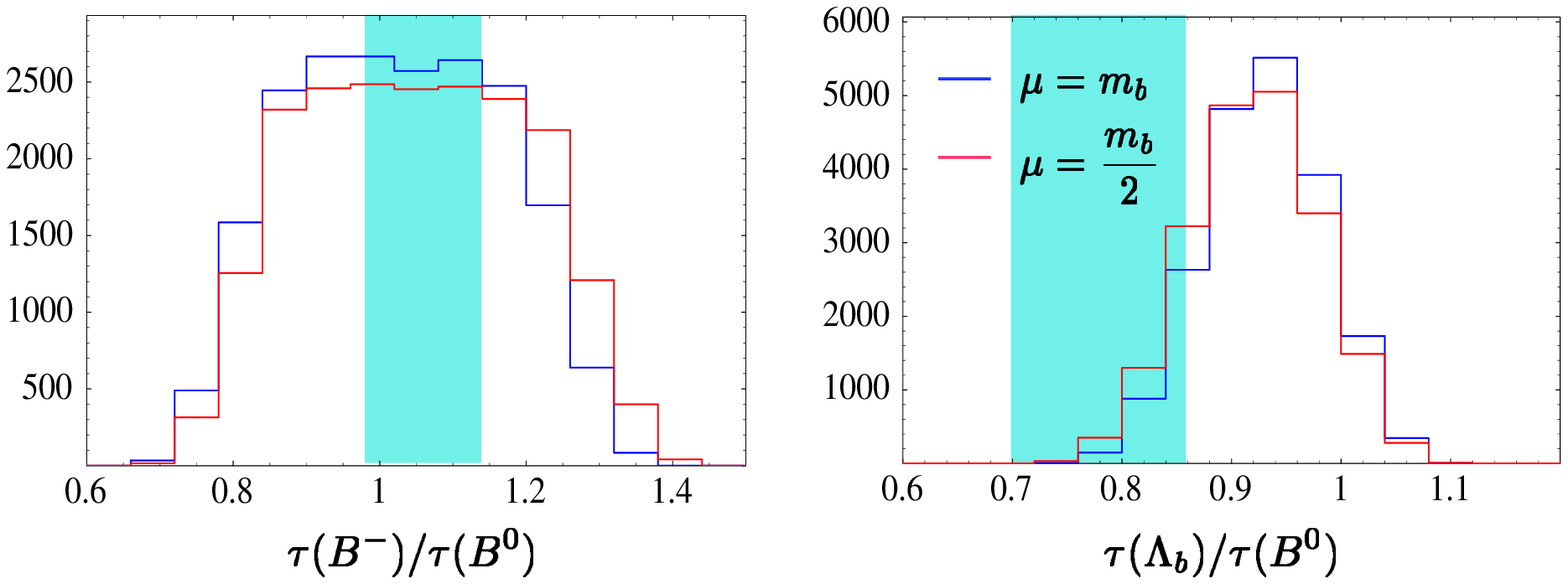}}
\begin{center}
\begin{minipage}[t]{6.3cm}
\begin{center}
\epsfxsize=5.5cm\epsffile{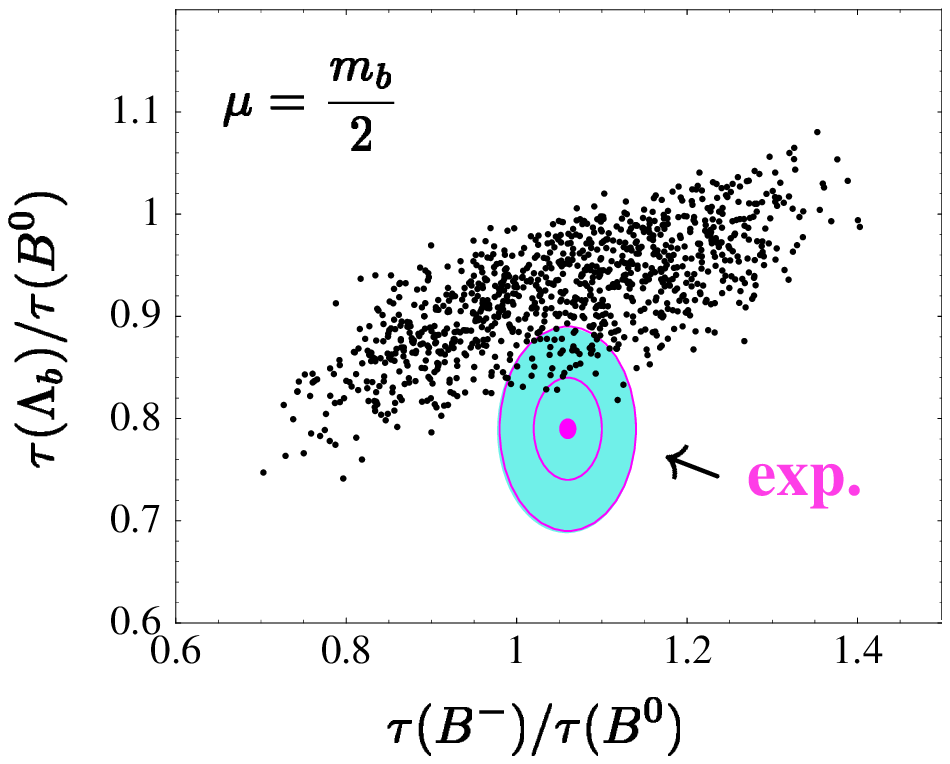}
\end{center}
\end{minipage}
\begin{minipage}[t]{5.3cm}
\begin{center}
\epsfxsize=5cm\epsffile{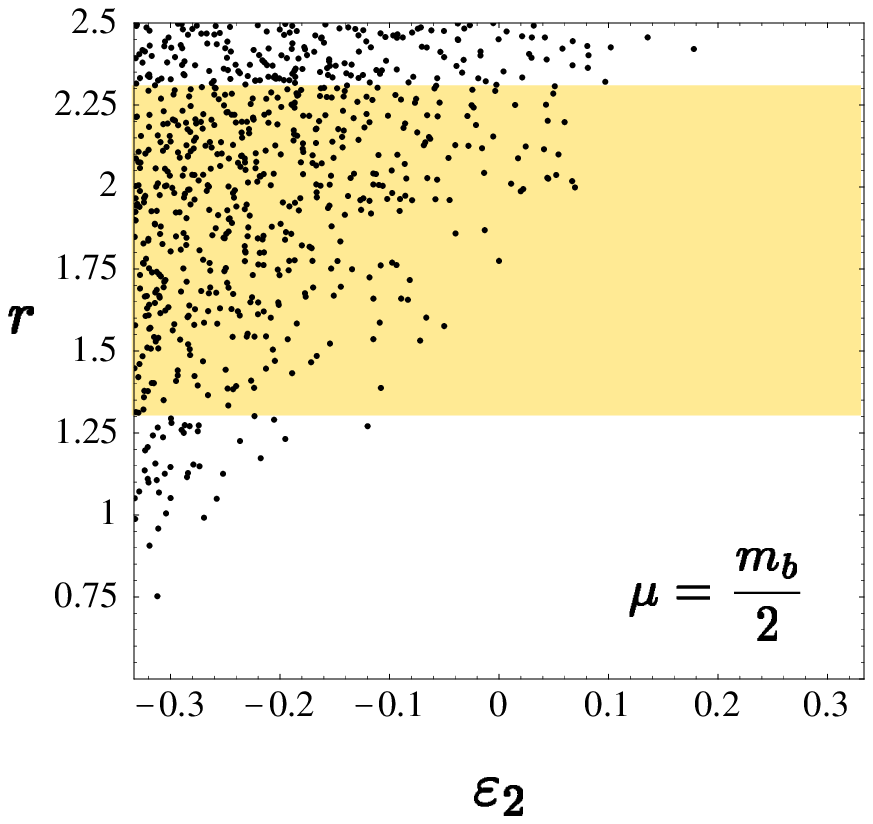}
\end{center}
\end{minipage}
\end{center}
\vspace{-0.5cm}
\caption{
Upper plots: Theoretical distributions for beauty lifetime ratios. The
two curves in each figure correspond to different choices of the
renormalization scale. The shaded areas show the experimental values
with their $2\sigma$ error bands. Lower left: Two-dimensional
distribution of lifetime ratios, together with the $1\sigma$ and
$2\sigma$ contours around the central experimental values. Lower right:
Distribution of the hadronic parameters $\varepsilon_2$ and $r$ for
simulations with results inside the $2\sigma$ ellipse.}
\label{fig:9to12}
\end{figure}

Without a reliable field-theoretical calculation of the bag parameters,
no lifetime ``predictions'' can be obtained. However, one can see which
ranges of the lifetime ratios can be covered using sensible values for
the bag parameters \cite{MNHawaii}. To this end, I scan the following
parameter space: $B_i,\widetilde B\in [2/3,4/3]$, $\varepsilon_i\in
[-1/3,1/3]$, $r\in [0.25,2.5]$, $m_c/m_b=0.29\pm 0.03$, $f_B=(200\pm
20)\,$MeV. The resulting distributions for the lifetime ratios are
shown in Fig.~\ref{fig:9to12}. The most important observation of this
exercise is that it is possible to reproduce both lifetime ratios,
within their experimental uncertainties, using the same set of input
parameters. The theoretically allowed range for the ratio
$\tau(B^-)/\tau(B^0)$ is centered around the experimental value;
however, the width of the allowed region is so large that no accurate
``prediction'' of this ratio could have been made. Any value between
0.8 and 1.3 could be easily accommodated by theory. The predictions for
the ratio $\tau(\Lambda_b)/\tau(B^0)$, on the other hand, center around
a value of 0.95, and the spread of the results is much narrower. Only a
small tail extends into the region preferred by experiment. Requiring
that the theoretical results be inside the $2\sigma$ ellipse around the
central experimental values, I find that the value of the parameter
$\varepsilon_1 $ must be close to zero, $\varepsilon_1\approx-(0.1\pm
0.1)$, in agreement with the expectation based on factorization. In
addition, this requirement maps out the region of parameter space for
$\varepsilon_2$ and $r$ shown in the right lower plot of
Fig.~\ref{fig:9to12}. Thus, the data indicate a large (with respect to
most model predictions) value of $r$ and a negative value of
$\varepsilon_2$. Ultimately, it will be important to perform reliable
field-theoretical calculations of the hadronic parameters, for instance
using lattice gauge theory or QCD sum rules, to see whether indeed the
observed beauty lifetime ratios can be accounted for by the heavy-quark
expansion (for a recent sum-rule estimate of the parameters $B_i$ and
$\varepsilon_i$, see Ref.~\cite{Baek}). This is particularly important
for the parameter $r$, large values of which are not excluded a priori.
Using a variant of the quark-model relation derived in
Ref.~\cite{Rosn96}, combined with a preliminary DELPHI measurement of
the mass-splitting between the $\Sigma_b^*$ and $\Sigma_b$ baryons
\cite{DELSigmab}, I find
\[
   r\approx \frac43\,\frac{M_{\Sigma_b^*}^2-M_{\Sigma_b}^2}
    {M_{B^*}^2-M_B^2} = 1.8\pm 0.5 \,.
\]
With such a large value, it is possible to explain the short
$\Lambda_b$ lifetime without invoking violations of local quark--hadron
duality.

Although at present there is thus no convincing evidence that the low
value of the $\Lambda_b$ lifetime could not be accommodated in the
context of the heavy-quark expansion, it remains a possibility that in
the future such a discrepancy may emerge, for instance if lattice
calculations would show that $r\ll 1$. Then, one would have to blame
violations of local quark--hadron duality to be responsible for the
failure of the heavy-quark expansion for inclusive, nonleptonic
$B$-decay rates. Recently, several authors have studied in QCD-inspired
models how quark--hadron duality may be violated. Based on a simple
model for the difference of two heavy--light current correlators in the
chiral and large-$N_c$ limits, Shifman has argued that deviations from
local duality are due to the asymptotic nature of the operator product
expansion (OPE) \cite{Shif94}. More recently, Blok et al.\ have studied
QCD$_2$ (i.e.\ QCD in $1+1$ space--time dimensions) at next-to-leading
order in $1/N_c$, where finite resonance widths provide for a dynamical
smearing of correlation functions \cite{Blok97}. They find that, in the
Euclidean region, deviations from local duality are indeed due to the
divergence of the OPE, and are exponentially suppressed with $Q^2$. In
the physical (Minkowskian) region, on the other hand, they find that
two mass scales, $\Lambda_1$ and $\Lambda_2$, are dynamically
generated. For $\Lambda_1^2<q^2<\Lambda_2^2$, there are strong
oscillations of correlation functions. Global duality works, but local
duality may be strongly violated. For $q^2>\Lambda_2$, the oscillations
become exponentially damped, and local duality is restored. They
conjecture that a similar pattern may hold for real, four-dimensional
QCD. Similar results have also been obtained by Grinstein and Lebed
\cite{GrinLeb}, and by Chibisov et al.\ studying the instanton vacuum
model \cite{Chib96}. Quite different conclusions have been reached by
Colangelo et al., who have considered a more ``realistic'' version of
Shifman's model, finding that global duality holds, but local duality
is spoiled by $1/Q$ effects not present in the OPE \cite{CDN97}. This
study gives some support for the conjecture by Altarelli et al.\ that
there may be non-OPE terms of order $\bar\Lambda/m_Q$ increasing the
nonleptonic decay rates of heavy hadrons \cite{AMPR}. Such terms could
easily explain the short $\Lambda_b$ lifetime. However, I stress that,
although the model of Ref.~\cite{CDN97} is interesting, it must not be
taken as an existence proof of non-OPE power corrections. Nature may be
careful enough to avoid power-like deviations from local duality.

\boldmath
\section{Hadronic $B$ decays}
\unboldmath

The strong interaction effects in hadronic decays are much more
difficult to understand, even at a qualitative level, than in leptonic
or semileptonic decays. The problem is that multiple gluon exchanges
can redistribute the quarks in the final state of a hadronic decay. As
a consequence, some phenomenological assumptions are unavoidable is
trying to understand these processes.

\subsection{Energetic two-body decays}

It has been argued that energetic two-body decays with a large energy
release are easier to understand because of the colour-transparency
phenomenon: a pair of fast-moving quarks in a colour-singlet state,
which is produced in a local interaction, effectively decouples from
long-wavelength gluons \cite{Bj89}. This intuitive argument suggests
that nonleptonic amplitudes factorize to a good approximation. The main
strong-interaction effects are then of a short-distance nature and
simply renormalize the operators in the effective weak Hamiltonian
\cite{Heff}. For the case of $b\to c\bar u d$ transitions, for
instance,
\[
   H_{\rm eff} = \frac{G_F}{\sqrt 2}\,V_{cb} V_{ud}^*\,\Big\{
   c_1(\mu)\,(\bar du) (\bar cb) + c_2(\mu)\,(\bar cu) (\bar db)
   \Big\} + \dots \,,
\]
where $(\bar d u)=\bar d\gamma^\mu(1-\gamma_5) u$ etc.\ are
left-handed, colour-singlet quark currents, and $c_1(m_b)\approx 1.1$
and $c_2(m_b)\approx -0.3$ are Wilson coefficients taking into account
the short-distance corrections arising from the exchange of hard
gluons. The effects of soft gluons remain in the hadronic matrix
elements of the local four-quark operators. In general, a reliable
field-theoretical calculation of these matrix elements is the obstacle
to a quantitative theory of hadronic weak decays.

Using Fierz identities, the four-quark operators in the effective
Hamiltonian may be rewritten in various forms. It is particularly
convenient to rearrange them in such a way that the flavour quantum
numbers of one of the quark currents match those of one of the hadrons
in the final state of the considered decay process. As an example,
consider the decays $\bar B\to D\pi$. Omitting common factors, the
various amplitudes can be written as
\begin{eqnarray}
   A_{\bar B^0\to D^+\pi^-} &=& \Big( c_1 + \frac{c_2}{N_c} \Big)
    \langle D^+\pi^-|(\bar d u)(\bar c b)|\bar B^0\rangle
    + 2 c_2\,\langle D^+\pi^-|(\bar d t_a u)(\bar c t_a b)
   |\bar B^0\rangle \nonumber\\
   &\equiv& a_1\,\langle\pi^-|(\bar d u)|0\rangle\,
    \langle D^+|(\bar c b)|\bar B^0\rangle\,, \nonumber\\
   A_{\bar B^0\to D^0\pi^0} &=& \Big( c_2 + \frac{c_1}{N_c} \Big)
    \langle D^0\pi^0|(\bar c u)(\bar d b)|\bar B^0\rangle
    + 2 c_1\,\langle D^0\pi^0|(\bar c t_a u)(\bar d t_a b)
   |\bar B^0\rangle \nonumber\\
   &\equiv& a_2\,\langle D^0|(\bar c u)|0\rangle\,
    \langle\pi^0|(\bar d b)|\bar B^0\rangle \,, \nonumber
\end{eqnarray}
where $t_a$ are the SU(3) colour matrices. The two classes of decays
shown above are referred to as class-1 and class-2, respectively. The
factorized matrix elements in the last steps are known in terms of the
meson decay constants $f_\pi$ and $f_D$, and the transition form
factors for the decays $\bar B\to D$ and $\bar B\to\pi$. Most of these
quantities are accessible experimentally. Of course, the above matrix
elements also contain other, nonfactorizable contributions. They are
absorbed into the definition of the hadronic parameters $a_1$ and
$a_2$, which in general are process dependent. Recently, some progress
in the understanding of these parameters has been made, leading to the
``generalized factorization hypothesis'' that
\cite{Chen}--\cite{Stech97}
\[
  a_1\approx c_1(m_b) \,,\qquad
  a_2\approx c_2(m_b) + \zeta c_1(m_b) \,,
\]
where $\zeta$ is a process-independent hadronic parameter (for
energetic two-body decays only!), which accounts for the dominant
nonfactorizable contributions to the decay amplitudes. To derive these
results, one combines the $1/N_c$ expansion with an argument inspired
by colour transparency \cite{Stech97}.

\begin{table}[t]
%\vspace{-0.5cm}
\caption{Experimental tests of the generalized factorization
hypothesis}
\begin{center}
\begin{tabular}{|lcr|}\hline
 & $a_1$ & $a_2/a_1~~~$ \\
\hline
{}~$\bar B\to D^{(*)}\pi$ & ~$1.08\pm 0.06$~ & ~$0.21\pm 0.07$~ \\
{}~$\bar B\to D^{(*)}\rho$ & ~$1.07\mp 0.07$~ & ~$0.23\pm 0.14$~ \\
{}~$\bar B\to\psi^{(\prime)} K^{(*)}$~ & & $|a_2|=0.21\pm 0.04$~ \\
{}~$\bar B\to D^{(*)}\bar D_s^{(*)}$~ & ~$1.10\pm 0.18$~ & \\
\hline
\end{tabular}
\end{center}
\vspace{-0.3cm}
\label{tab:4}
\end{table}

Some experimental tests of the generalized factorization hypothesis are
shown in Table~\ref{tab:4}, where I quote the values of $a_1$ and $a_2$
extracted from the analysis of different classes of decay modes, using
data reported by CLEO \cite{CLEOhad}--\cite{Miller}. Within the present
experimental errors, there is indeed no evidence for any process
dependence of the hadronic parameters. The generalized factorization
prescription provides a simultaneous description of all measured
Cabibbo-allowed two-body decays with a single parameter $\zeta =
0.45\pm 0.05$ extracted from the data on class-2 decays \cite{Stech97}.
So far, this theoretical framework is fully supported by the data.

The factorization hypothesis can be used to obtain rather precise
values for the decay constants of the $D_s$ and $D_s^*$ mesons, by
comparing the theoretical predictions for various ratios of $\bar B^0$
decay rates,
\begin{eqnarray}
   \frac{\Gamma(D^+ D_s^-)}{\Gamma(D^+\pi^-)}
   &=& 1.01\,\left( \frac{f_{D_s}}{f_\pi} \right)^2
    \,, \hspace{0.87cm}
   \frac{\Gamma(D^{*+} D_s^-)}{\Gamma(D^{*+}\pi^-)}
   = 0.72\,\left( \frac{f_{D_s}}{f_\pi} \right)^2
    \,, \nonumber\\
   \frac{\Gamma(D^+ D_s^{*-})}{\Gamma(D^+\rho^-)}
   &=& 0.74\,\left( \frac{f_{D_s^*}}{f_\rho} \right)^2
    \,, \qquad
   \frac{\Gamma(D^{*+} D_s^{*-})}{\Gamma(D^{*+}\rho^-)}
   = 1.68\,\left( \frac{f_{D_s^*}}{f_\rho} \right)^2 \,, \nonumber
\end{eqnarray}
with data. These predictions are rather clean for the following
reasons: first, all decays involve class-1 transitions, so that
deviations from factorization are probably very small; secondly, the
parameter $a_1$ cancels in the ratios; thirdly, the two processes in
each ratio have a similar kinematics, so that the corresponding decay
rates are sensitive to the same form factors, however evaluated at
different $q^2$ values. Finally, some of the experimental systematic
errors cancel in the ratios (however, I do not assume this in quoting
errors below). Combining these predictions with the average
experimental branching ratios \cite{CLEOhad} yields
\[
   f_{D_s} = (234\pm 25)~{\rm MeV} \,, \qquad
   f_{D_s^*} = (271\pm 33)~{\rm MeV} \,.
\]
The result for $f_{D_s}$ is in good agreement with the value
$f_{D_s}=250\pm 37$~MeV extracted from the leptonic decay
$D_s\to\mu^+\nu$ \cite{Besson}. The ratio of decay constants,
$f_{D_s^*}/f_{D_s}=1.16\pm 0.19$, which cannot be determined from
leptonic decays, is in good agreement with theoretical expectations
\cite{Nari94,Huan94}. Finally, I note that, assuming SU(3)-breaking
effects of order 10--20\%, the established value of $f_{D_s}$ implies
$f_D\gsim 200$~MeV, which is larger than most theoretical predictions.

\boldmath
\subsection{$B$ Decays into two light mesons}
\unboldmath

\begin{figure}
\vspace{-0.5cm}
\epsfxsize=8cm
\centerline{\epsffile{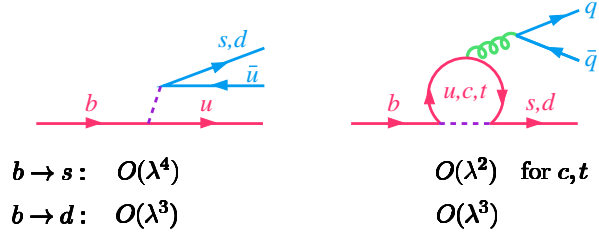}}
%\vspace{-0.5cm}
\caption{\label{fig:14}
Tree and penguin diagrams for rare $B$ decays into light hadrons}
\end{figure}

Since last year, CLEO has observed (or put strict upper limits on) a
number of rare $B$ decay modes, many of which have strongly suppressed
tree amplitudes and are thus dominated by loop processes. The relevant
quark diagrams for the transitions $b\to s(d)\bar qq$ are shown in
Fig.~\ref{fig:14}, where I also give the powers of the Wolfenstein
parameter $\lambda\approx 0.22$ associated with these diagrams. Because
the penguin digram shown on the right is a loop process, it is
sensitive to new heavy particles and thus potentially probes physics
beyond the Standard Model.

Some of the experimental results that have caused a lot of excitement
are (all in units of $10^{-5}$ and averaged over CP-conjugate modes;
upper limits at 90\% CL) \cite{Miller,CLEpiK,CLEeta}:
\begin{eqnarray}
   {\rm B}(B^0\to K^+\pi^-) &=& 1.5_{-0.4}^{+0.5}\pm 0.1\pm 0.1 \,,
    \nonumber\\
   {\rm B}(B^+\to K^0\pi^+) &=& 2.3_{-1.0}^{+1.1}\pm 0.3\pm 0.2 \,,
    \nonumber\\
   {\rm B}(B^0\to\pi^+\pi^-) &=& 0.7\pm 0.4 < 1.5 \,, \nonumber\\
   {\rm B}(B^+\to\pi^+\pi^0) &=& 1.0_{-0.5}^{+0.6} < 2.0 \,,
    \nonumber\\
   {\rm B}(B^+\to K^+\eta') &=& 7.1_{-2.1}^{+2.5}\pm 0.9 \,,
    \nonumber\\
   {\rm B}(B\to\eta'+X_s) &=& 62\pm 16\pm 13 \quad
    (2.0 < p_{\eta'}~[{\rm GeV}] < 2.7) \,. \nonumber
\end{eqnarray}
In Table~\ref{tab:5}, I show the results of an analysis of these
results in terms of SU(3)-invariant amplitudes representing diagrams
having tree topology $T^{(\prime)}$, colour-suppressed tree topology
$C^{(\prime)}$, penguin topology $P^{(\prime)}$, and SU(3)-singlet
penguin topology $S^{(\prime)}$, where primed (unprimed) quantities
refer to $b\to s(d)$ transitions \cite{Dighe97}. The results of this
analysis establish the presence of a significant SU(3)-singlet
contribution $S'$ active in the decays with an $\eta'$ meson in the
final state. They also indicate a rather large ``penguin pollution''
$|P/T|\approx 0.3\pm 0.1$ in the decays $B\to\pi\pi$. This has
important (and optimistic) implications for CP-violation studies
using these decay modes \cite{CPpiK}.

\begin{table}
\vspace{-0.5cm}
\caption{\label{tab:5}
Amplitude analysis for some rare $B$ decays into two light mesons. The
dominant amplitudes are highlighted.}
\vspace{-0.15cm}
\epsfxsize=10cm
\centerline{\epsffile{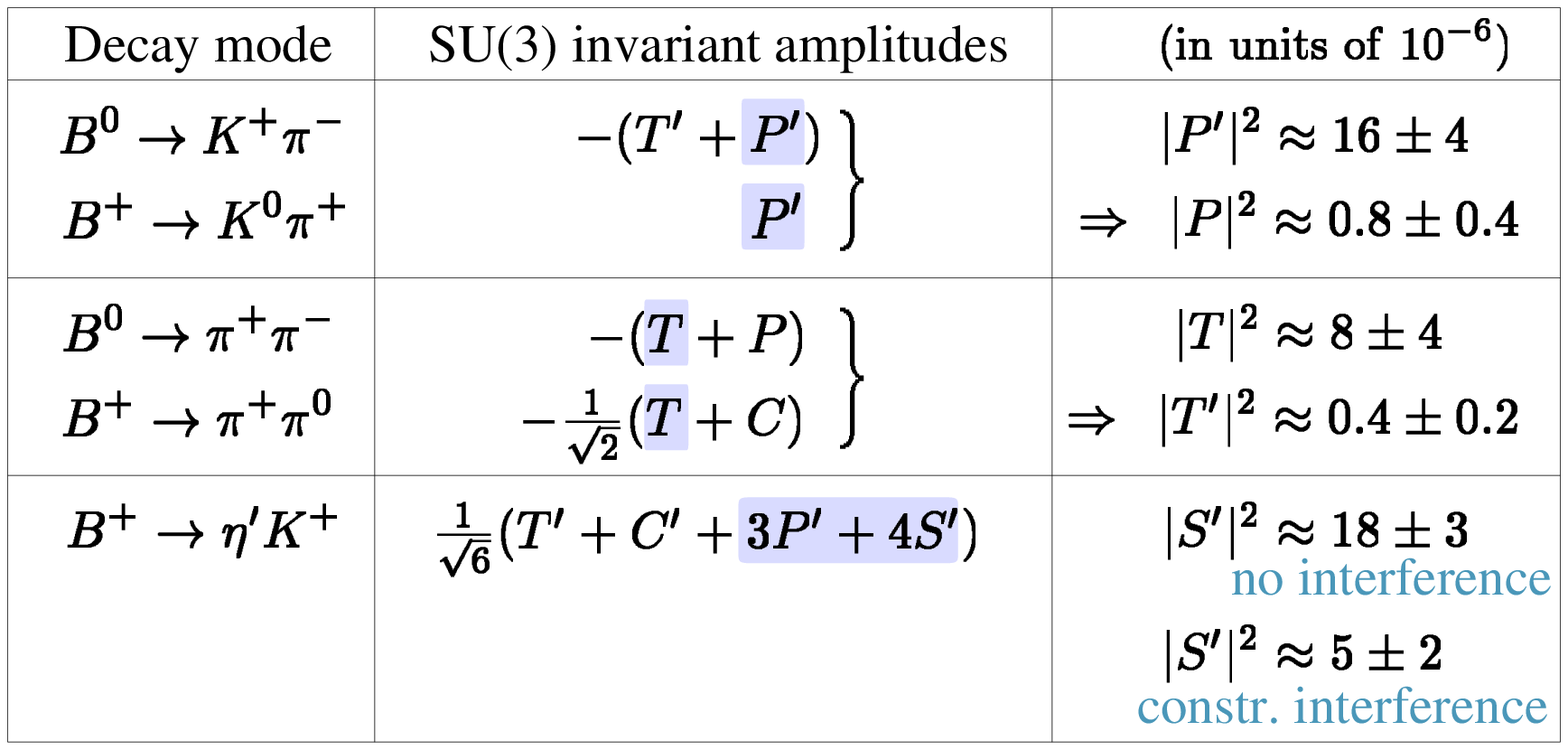}}
\vspace{-0.3cm}
\end{table}

\begin{figure}
\vspace{-0.5cm}
\epsfxsize=8cm
\centerline{\epsffile{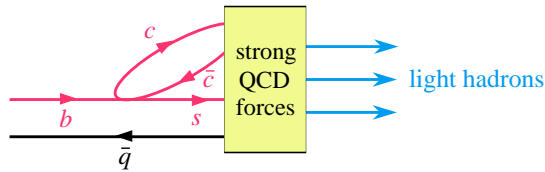}}
\vspace{-0.5cm}
\caption{\label{fig:15}
Long-distance effects in ``charming penguins''}
\end{figure}

Several authors have discussed the question whether the penguin
amplitudes in these transitions are anomalously large. There seems to
be some agreement that a standard analysis, using next-to-leading order
Wilson coefficients combined with factorized matrix elements, can
accommodate the experimental results for the exclusive rare decay rates
\cite{AliGreub,Desh97}. However, it has been stressed that significant
long-distance contributions to the charm penguin are required to fit
the data \cite{BuFl,charmp}. This is not too surprising, since the
energy release in $B$ decays is such that the $c\bar c$ pair in the
charm penguin is not far from its mass shell, and hence this penguin is
really more a long-distance than a short-distance process
\cite{BjHawaii}. This is illustrated in Fig.~\ref{fig:15}. In the case
of $B\to K\pi$ decays, the box may represent final-state rescattering
processes such as $B\to D_s\bar D\to K\pi$ \cite{resc1}--\cite{resc4}.
If the final state contains an $\eta'$ meson, the box may represent an
anomaly-mediated coupling of the $\eta'$ to glue, or an intrinsic charm
component in the $\eta'$ wave function. At least to some extent these
are different words for the same physics.

\begin{figure}
\vspace{-0.5cm}
\epsfxsize=10cm
\centerline{\epsffile{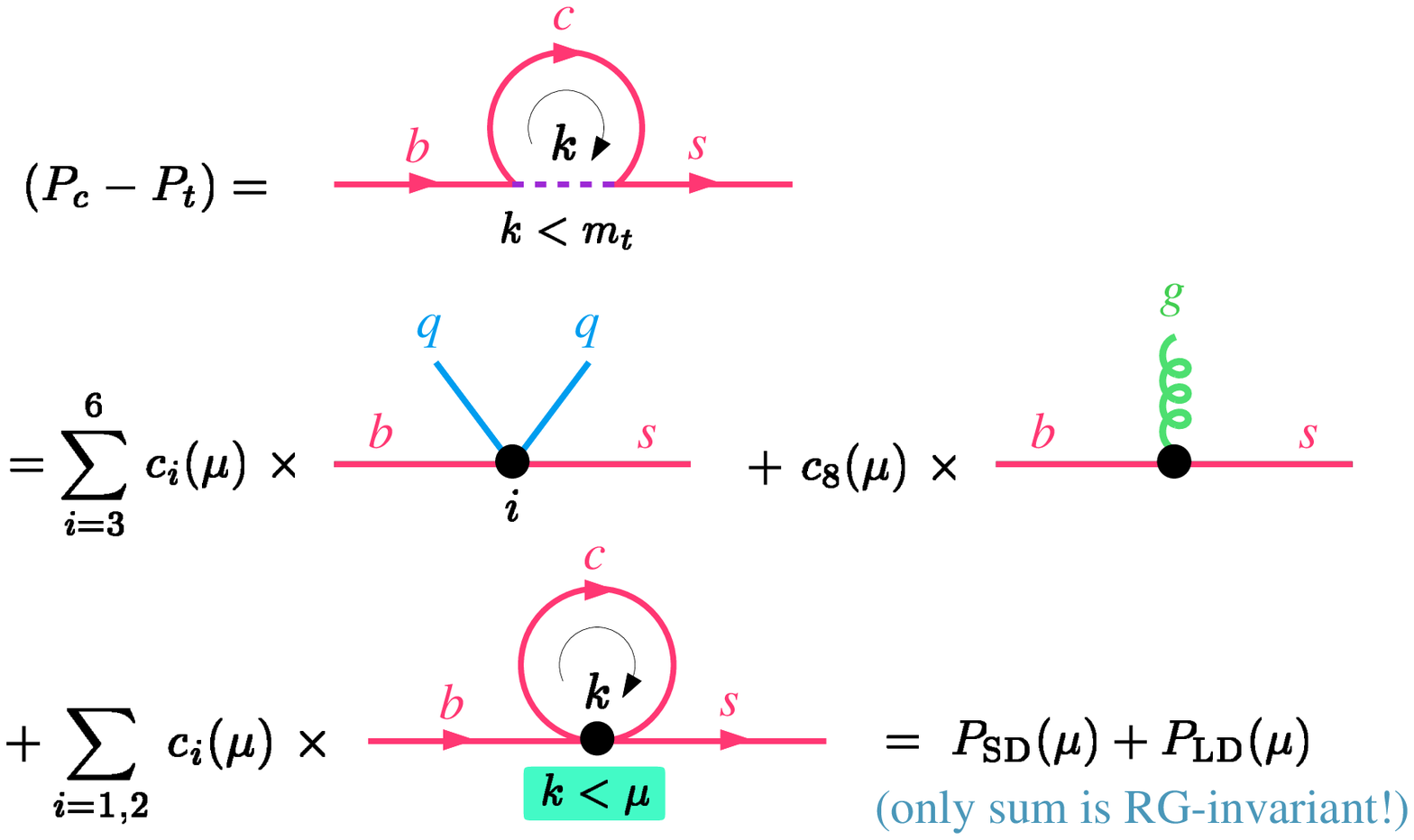}}
\vspace{-0.2cm}
\caption{\label{fig:16}
Operator product expansion for the charm penguin}
\end{figure}

In a way, the charm penguin is all there is in $b\to s$ FCNC processes
\cite{BjHawaii}, because the unitarity of the CKM matrix implies that
\[
   \sum_{q=u,c,t} V_{qs}^* V_{qb} P_q
   \approx V_{cs}^* V_{cb} (P_c - P_t) \,.
\]
The top penguin $P_t$ simply provides the GIM cutoff for large loop
momenta, whereas the up penguin $P_u$ is strongly CKM suppressed. The
construction of the effective weak Hamiltonian is used to separate the
short- and long-distance contributions to the charm penguin, in a way
that is illustrated in Fig.~\ref{fig:16}. The short-distance
contributions from large loop momenta are contained in the local
operators $Q_{3,\dots,6}$ and $Q_8$, while long-distance effects from
small momenta are contained in the matrix elements of the
current--current operators $Q_{1,2}$. From the physical argument
presented above, one expects sizable long-distance contributions, since
the energy release in $B$ decays is in the region of charm resonances.
Therefore, purely short-distance estimates of the charm penguin may be
very much misleading.

\boldmath
\subsection{Explanations of the $\eta'$ modes}
\unboldmath

To explain the yield of $\eta'$ mesons in rare $B$ decays, in
particular the large inclusive $\eta'$ production rate, is a challenge
to theorists. Currently, the situation is still controversial, and
several proposals are being discussed. It has been suggested that the
process $b\to s g^*$ followed by $g^*\to\eta' g$ is enhanced, either by
the anomalous coupling of the $\eta'$ to glue
\cite{AtSoni}--\cite{Frit}, or because the initial $b\to sg$ rate is
enhanced by some new physics \cite{KaPetr}. Because of its three-body
nature, this mechanism would mainly be responsible for multi-particle
inclusive processes containing $\eta'$ mesons, but give a minor
contribution to the two-body modes $B\to\eta' K^{(*)}$. It has also
been argued that an enhanced $\eta'$ production could result from a
significant intrinsic charm component in the $\eta'$ wave function,
either through the colour-singlet mechanism $b\to(c\bar c)_1\,s$
followed by $(c\bar c)_1\to\eta'$ \cite{Halp,Shur97}, or through the
colour-octet mechanism $b\to(c\bar c)_8\,s$ followed by $(c\bar
c)_8\to\eta' X$ \cite{Yuan}, which is strongly favoured by the
structure of the effective weak Hamiltonian. The relevance of the
intrinsic charm mechanism depends on the size of a ``decay constant''
$f_{\eta'}^{(c)}$, estimates of which range from 6~MeV to 180~MeV. A
clarification of this issue is needed (see, e.g., the discussion in
Refs.~\cite{AliGreub,Petrov}). Yet other possibilities to explain the
large $\eta'$ yield have been explored in Refs.~\cite{Ahmady,DuKim97}.

\begin{table}
\vspace{-0.5cm}
\caption{Model predictions for ratios of rare $B$ decay rates}
\vspace{-0.5cm}
\begin{center}
\begin{tabular}{|l|ccccc|}\hline
 & ~Ali, Greub~ & ~Lipkin~ & ~Cheng, Tseng~ & ~Datta et al.~ &
 ~Halperin et al.~ \\
 & \protect\cite{AliGreub} & \protect\cite{Lipkin} &
 \protect\cite{Cheng97} & \protect\cite{Datta} &
 \protect\cite{Halp} \\
\hline
$\displaystyle\frac{B^+\to\eta' K^{*+}}{B^+\to\eta' K^+}$ &
 0.01--0.2 & & 0.20--0.24 & 0.36--0.38 & $\approx 2$ \\
$\displaystyle\frac{B^+\to\eta K^+}{B^+\to\eta' K^+}$ &
 0.04--0.14 & 0--0.02 & 0.03--0.08 & 0.02--0.09 & \\
$\displaystyle\frac{B^+\to\eta K^{*+}}{B^+\to\eta' K^{*+}}$ &
 0.6--2.9 & 0.5--8.2 & 0.26--0.48 & 1.0--1.4 & \\
\hline
\end{tabular}
\end{center}
\vspace{-0.3cm}
\label{tab:6}
\end{table}

Future measurements of other decay channels, as well as of the
distribution of the invariant hadronic mass $M(X_s)$ in the decay
$B\to\eta' X_s$, will help to clarify the situation. Preliminary CLEO
data indicate that relatively large values of $M(X_s)$ are preferred
\cite{Miller,CLEMXs}. Another good discriminator between models is
provided by the ratios of the various $B\to\eta^{(\prime)} K^{(*)}$
decay rates. A collection of theoretical predictions for such ratios is
shown in Table~\ref{tab:6}. In summary, the rare decays of $B$ mesons
into light mesons provide unique opportunities to extract many yet
unknown hadronic matrix elements. At present, there are no convincing
arguments that these processes could not be understood in a
conventional Standard Model framework. In particular, because of the
complexity of the dynamics of the $\eta'$ meson, I believe it is
premature to deduce evidence for new physics from the yield of $\eta'$
production in $B$ decays.

\section{Summary and outlook}

Owing to the combined efforts of experimenters and theorists, there has
been significant progress in heavy-quark physics in the past few years.
The exclusive and inclusive semileptonic decays mediated by the
transition $b\to c\,\ell\,\bar\nu$ are understood from first
principles, using the heavy-quark expansion. Heavy-quark effective
theory (HQET) works well in describing these processes, and starts
being tested at the level of symmetry-breaking corrections. This, by
itself, is a remarkable development, which puts the determination of
$|V_{cb}|=0.038\pm 0.003$ on firm theoretical grounds. New ideas are
being developed for getting a precise $(\sim 10\%$) determination of
$|V_{ub}|$, combining exclusive and inclusive methods. The error in the
current value, $|V_{ub}|=(3.3\pm 0.8)\times 10^{-3}$, is still
dominated by the theoretical uncertainty.

New flavour-specific measurements of semi-inclusive charm yields
indicate that the ``problem'' of the charm deficit and low semileptonic
branching ratio is disappearing. The theoretical predictions for
$\mbox{B}(\bar B\to X_{c\bar c})$ and $\mbox{B}(\bar B\to
\mbox{no~charm})$ are confirmed by the data. There is no compelling
evidence for violations of quark--hadron duality, nor for new physics
or exotic decay modes in $B$ decays. The low value of the $\Lambda_b$
lifetime remains a surprise, but can be accommodated in a small corner
of parameter space. Lattice calculations of baryon matrix elements of
four-quark operators would help to clarify the situation. However, this
may also be the first sign of departures from local quark--hadron
duality.

Hadronic two-body decays of $B$ mesons with a large energy release can
be understood in terms of a generalized factorization hypothesis, which
includes the leading nonfactorizable corrections. Finally, new data on
$B$ decays into two light mesons open a window to study the details of
nonperturbative dynamics of QCD, including long-distance penguins and
other exotica. So far, the Standard Model can account for the data, but
we should be prepared to find surprises as measurements and theory
become more precise. For now, it remains to stress that the observed
penguin effects are large and thus raise hopes for having large CP
asymmetries in these decays. Hence, there is more $B$eauty to come in a
bright future.


\begin{thebibliography}{999}

\bibitem {review}
For a review and a comprehensive list of references before 1996, see:
M.~Neubert, Phys.\ Rep.\ {\bf 245}, 259 (1994);
Int.\ J.\ Mod.\ Phys.\ A {\bf 11}, 4173 (1996).

\bibitem {ABN97}
G. Amor\'os, M. Beneke and M. Neubert, Phys.\ Lett.\ B {\bf 401}, 81
(1997).

\bibitem {CzGr97}
A. Czarnecki and A.G. Grozin, Phys.\ Lett.\ B {\bf 405}, 142 (1997).

\bibitem {GrNe97}
A.G. Grozin and M. Neubert, Nucl.\ Phys.\ B {\bf 508}, 311 (1997).

\bibitem {FLS2}
A.F. Falk, M. Luke and M.J. Savage, Phys.\ Rev.\ D {\bf 53}, 6316
(1996).

\bibitem {GKLW}
M. Gremm et al. Phys.\ Rev.\ Lett.\ {\bf 77}, 20 (1996).
%M. Gremm, A. Kapustin, Z. Ligeti and M.B. Wise, Phys.\ Rev.\ Lett.\
%{\bf 77}, 20 (1996).

\bibitem {Cher}
V. Chernyak, Nucl.\ Phys.\ B {\bf 457}, 96 (1995);
Phys.\ Lett.\ B {\bf 387}, 173 (1996).

\bibitem {GrSt96}
M. Gremm and I. Stewart, Phys.\ Rev.\ D {\bf 55}, 1226 (1997).

\bibitem {LiYu}
H. Li and H.-L. Yu, Phys.\ Rev.\ D {\bf 55}, 2833 (1997).

\bibitem {Vcb}
M. Neubert, Phys.\ Lett.\ B {\bf 264}, 455 (1991); {\bf 338}, 84
(1994).

\bibitem {BaBar}
The theoretical error in the prediction for ${\cal F}(1)$ has been
reanalysed in {\sl The {\sc BaBar} Physics Books}, Chapter~8, to appear
as a SLAC Report.

\bibitem {Cz97}
A. Czarnecki, Phys.\ Rev.\ Lett.\ {\bf 76}, 4124 (1996);
A.~Czarnecki and K.~Melnikov, Nucl.\ Phys.\ B {\bf 505}, 65 (1997).

\bibitem {Franz}
J. Franzkowski and J.B. Tausk, Preprint MZ-TH/97-37 [hep-ph/9712205].

\bibitem {Luke}
M.E. Luke, Phys.\ Lett.\ B {\bf 252}, 447 (1990).

\bibitem {Ur97}
N. Uraltsev, Nucl.\ Phys.\ B {\bf 491}, 303 (1997);
A. Czarnecki, K. Melnikov and N. Uraltsev, Preprint TPI-MINN-97-19
[hep-ph/9706311].

\bibitem {BoydVcb}
C.G. Boyd, B. Grinstein and R.F. Lebed, Phys.\ Lett.\ B {\bf 353}, 306
(1995);
Nucl.\ Phys.\ B {\bf 461}, 493 (1996);
Phys.\ Rev.\ D {\bf 56}, 6895 (1997).

\bibitem {CN96}
I. Caprini and M. Neubert, Phys.\ Lett.\ B {\bf 380}, 376 (1996).

\bibitem {CLN97}
I. Caprini, L. Lellouch and M. Neubert, Preprint CERN-TH/97-91
[hep-ph/9712417].

\bibitem {Persis}
P. Drell, Preprint CLNS-97-1521 [hep-ex/9711020], to appear in the
Proceedings of the 18th International Symposium on Lepton--Photon
Interactions, Hamburg, Germany, July 1997.

\bibitem {GibbWars}
L.K. Gibbons, Proceedings of the 28th International Conference on High
Energy Physics, Warsaw, Poland, July 1996, edited by Z.~Ajduk
and A.K.~Wroblewski (World Scientific, Singapore, 1997), p.~183.

\bibitem {ALEPHGF}
ALEPH Coll.\ (D. Buskulic et al.), Phys.\ Lett.\ B {\bf 395}, 373
(1997).

\bibitem {CLEOGF}
CLEO Coll.\ (M. Athanas et al.), Phys.\ Rev.\ Lett.\ {\bf 79}, 2208
(1997).

\bibitem {CLEOtests}
CLEO Coll.\ (J.E. Duboscq et al.), Phys.\ Rev.\ Lett.\ {\bf 76}, 3898
(1996);
CLEO Coll.\ (A. Anastassov et al.), CLEO~CONF~96-8, contributed paper
to the 28th International Conference on High Energy Physics, Warsaw,
Poland, July 1996.

\bibitem {Leib}
A.K. Leibovich et al., Phys.\ Rev.\ Lett.\ {\bf 78}, 3995 (1997);
Phys.\ Rev.\ D {\bf 57}, 308 (1998).
%A.K. Leibovich, Z. Ligeti, I.W. Steward and M.B. Wise, Phys.\ Rev.\
%Lett.\ {\bf 78}, 3995 (1997); Phys.\ Rev.\ D {\bf 57}, 308 (1998).

\bibitem {MNpwave}
M. Neubert, Preprint CERN-TH/97-240 [hep-ph/9709327], to appear in
Phys.\ Lett.\ B.

\bibitem {CLEOVub}
CLEO Coll.\ (J.P. Alexander et al.), Phys.\ Rev.\ Lett.\ {\bf 77}, 5000
(1996).

\bibitem {lattVub}
UKQCD Coll.\ (J.M. Flynn et al.), Nucl.\ Phys.\ B {\bf 461}, 327
(1996);
for a review, see: J.M. Flynn and C.T. Sachrajda, Preprint SHEP-97-20
[hep-lat/9710057], to appear in {\sl Heavy Flavours}, Second Edition,
edited by A.J.~Buras and M.~Linder (World Scientific, Singapore).

\bibitem {SRVub}
A. Khodjamirian et al., Phys.\ Lett.~B {\bf 410}, 275 (1997).
%A. Khodjamirian, R. R\"uckl, S. Weinzierl and O. Yakovlev, Phys.\
%Lett.~B {\bf 410}, 275 (1997).

\bibitem {BB97}
P. Ball and V.M. Braun, Phys.\ Rev.\ D {\bf 55}, 5561 (1997);
E. Bagan, P.~Ball and V.M.\ Braun, Preprint NORDITA-97-59
[hep-ph/9709243].

\bibitem {dispVub}
L. Lellouch, Nucl.\ Phys.\ B {\bf 479}, 353 (1996).

\bibitem{BoydVub}
C.G. Boyd, B. Grinstein and R.F. Lebed, Phys.\ Rev.\ Lett.\ {\bf 74},
4603 (1995);
C.G. Boyd and M.J. Savage, Phys.\ Rev.\ D {\bf 56}, 303 (1997).

\bibitem {BurdVub}
G. Burdman and J. Kambor, Phys.\ Rev.\ D {\bf 55}, 2817 (1997).

\bibitem {Gibbons}
L.K. Gibbons, to appear in the Proceedings of the 7th International
Symposium on Heavy Flavour Physics, Santa Barbara, California, July
1997.

\bibitem{Barg90}
V. Barger, C.S. Kim and R.J.N. Phillips, Phys.\ Lett.\ B {\bf 251}, 629
(1990).

\bibitem{Dai94}
J. Dai, Phys.\ Lett.\ B {\bf 333}, 212 (1994).

\bibitem{Bouz94}
A.O. Bouzas and D. Zappala, Phys.\ Lett.\ B {\bf 333}, 215 (1994).

\bibitem {GrRe97}
C. Greub and S.-J. Rey, Phys.\ Rev.\ D {\bf 56}, 4250 (1997).

\bibitem {Dike97}
R.D. Dikeman and N.G. Uraltsev, Preprint TPI-MINN-97-06-T
[hep-ph/9703437];
I. Bigi, R.D. Dikeman and N. Uraltsev, Preprint TPI-MINN-97-21-T
[hep-ph/9706520].

\bibitem {FLW97}
A.F. Falk, Z. Ligeti and M.B. Wise, Phys.\ Lett.\ B {\bf 406}, 225
(1997).

\bibitem {CJin}
C. Jin, Preprint DO-TH-97-27 [hep-ph/9801230].

\bibitem {MN94}
M. Neubert, Phys.\ Rev.\ D {\bf 49}, 3392 and 4623 (1994).

\bibitem {BSUV94}
I.I. Bigi et al., Int.\ J.\ Mod.\ Phys.\ A {\bf 9}, 2467 (1994).
%I.I. Bigi, M.A. Shifman, N.G. Uraltsev and A.I. Vainshtein, Int.\ J.\
%Mod.\ Phys.\ A {\bf 9}, 2467 (1994).

\bibitem {Baga}
E. Bagan et al., Nucl.\ Phys.\ B {\bf 432}, 3 (1994);
Phys.\ Lett.\ B {\bf 342}, 362 (1995) [E: {\bf 374}, 363 (1996)];
E. Bagan et al., Phys.\ Lett.\ B {\bf 351}, 546 (1995).
%E. Bagan, P. Ball, V.M. Braun and P. Gosdzinsky, Nucl.\ Phys.\ B {\bf
%432}, 3 (1994); Phys.\ Lett.\ B {\bf 342}, 362 (1995) [E: {\bf 374},
%363 (1996)];
%E. Bagan, P. Ball, B. Fiol and P. Gosdzinsky, Phys.\ Lett.\ B {\bf
%351}, 546 (1995).

\bibitem {NS97}
M. Neubert and C.T. Sachrajda, Nucl.\ Phys.\ B {\bf 483}, 339 (1997).

\bibitem {Michael}
M. Feindt, these Proceedings.

\bibitem {Uli97}
A. Lenz, U. Nierste and G. Ostermaier, Phys.\ Rev.\ D {\bf 56}, 7228
(1997).

\bibitem {Buch95}
G. Buchalla, I. Dunietz and H. Yamamoto, Phys.\ Lett.\ B {\bf 364}, 188
(1995).

\bibitem {CLEOwrongc}
CLEO Coll.\ (T.E. Coan et al.), Preprint CLNS-97-1516 [hep-ex/9710028].

\bibitem {Du97}
I. Dunietz et al., Eur.\ Phys.\ J.~C {\bf 1}, 211 (1998).
%I. Dunietz, J. Incandela, F.D. Snider and H. Yamamoto, Eur.\ Phys.\
%J.~C {\bf 1}, 211 (1998).

\bibitem {Ra97}
A.L. Kagan and J. Rathsman, Preprint [hep-ph/9701300];
A.L. Kagan, these Proceedings.

\bibitem {Tom97}
T. Browder, to appear in the Proceedings of the 7th International
Symposium on Heavy Flavor Physics, Santa Barbara, California, July
1997.

\bibitem {Close97}
F.E. Close et al., Preprint RAL-97-036 [hep-ph/9708265].

\bibitem {Alex95}
A. Kagan, Phys.\ Rev.\ D {\bf 51}, 6196 (1995).

\bibitem {RoSh96}
L. Roszkowski and M. Shifman, Phys.\ Rev.\ D {\bf 53}, 404 (1996).

\bibitem {CLDDbar}
CLEO Coll., CLEO~CONF~97-26, contributed paper eps97-337 to this
Conference.

\bibitem {ALDDbar}
ALEPH Coll., contributed paper pa05-060 to the 28th International
Conference on High Energy Physics, Warsaw, Poland, July 1996.

\bibitem {DELDDbar}
DELPHI Coll., contributed paper pa01-108 to the 28th International
Conference on High Energy Physics, Warsaw, Poland, July 1996.

\bibitem {DELbccs}
DELPHI Coll., DELPHI~97-80 CONF~66, contributed paper eps97-448 to this
Conference;
P.~Kluit, these Proceedings.

\bibitem {SLDnew}
M. Douadi, these Proceedings.

\bibitem {lifetimes}
T. Junk, to appear in the Proceedings of the 2nd International
Conference on $B$ Physics and CP Violation, Honolulu, Hawaii, March
1997.

\bibitem {BBD96}
M. Beneke, G. Buchalla and I. Dunietz, Phys.\ Rev.\ D {\bf 54}, 4419
(1996).

\bibitem {Blok94}
B. Blok et al., Phys.\ Rev.\ D {\bf 49}, 3356 (1994) [E: {\bf 50},
3572 (1994)];
%B. Blok, L. Koyrakh, M.A. Shifman and A.I. Vainshtein, Phys.\ Rev.\
%D {\bf 49}, 3356 (1994) [E: {\bf 50}, 3572 (1994)];
I.I. Bigi et al., in: B Decays, edited by S. Stone, Second Edition
(World Scientific, Singapore, 1994), p.~134.

\bibitem {Rosn96}
J.L. Rosner, Phys.\ Lett.\ B {\bf 379}, 267 (1996).

\bibitem {Cola96}
P. Colangelo and F. De Fazio, Phys.\ Lett.\ B {\bf 387}, 371 (1996).

\bibitem {MNHawaii}
M. Neubert, Preprint CERN-TH/97-148 [hep-ph/9707217], to appear in the
Proceedings of the 2nd International Conference on $B$ Physics and CP
Violation, Honolulu, Hawaii, March 1997.

\bibitem {Baek}
M.S. Baek, J. Lee, C. Liu and H.S. Song, Preprint SNUTP-97-064
[hep-ph/9709386].

\bibitem {DELSigmab}
DELPHI Coll.\ (P. Abreu et al.), DELPHI 95-107 PHYS 542, contributed
paper to the 28th International Conference on High Energy Physics,
Warsaw, Poland, July 1996.

\bibitem {Shif94}
M. Shifman, in {\sl Particles, Strings and Cosmology}, Proceedings of
the Joint Meeting of the International Symposium on Particles, Strings
and Cosmology and the 19th Johns Hopkins Workshop on Current Problems
in Particle Theory, edited by J.~Bagger et al.\ (World Scientific,
Singapore, 1996), p.~69.

\bibitem {Blok97}
B. Blok, M. Shifman and D.-X. Zhang, Preprint TPI-MINN-13-97-T
[hep-ph/9709333].

\bibitem {GrinLeb}
B. Grinstein and R.F. Lebed, Preprint UCSD/PTH~97-20 [hep-ph/9708396].

\bibitem {Chib96}
B. Chibisov et al., Int.\ J.\ Mod.\ Phys.\ A {\bf 12}, 2075 (1997).
%B. Chibisov, R.D. Dikeman, M. Shifman and N. Uraltsev, Int.\ J.\ Mod.\
%Phys.\ A {\bf 12}, 2075 (1997).

\bibitem {CDN97}
P. Colangelo, C.A. Dominguez and G. Nardulli, Phys.\ Lett.\ B {\bf
409}, 417 (1997).

\bibitem {AMPR}
G. Altarelli et al., Phys.\ Lett.\ B {\bf 382}, 409 (1996).
%G. Altarelli, G. Martinelli, S. Petrarca and F. Rapuano, Phys.\ Lett.\
%B {\bf 382}, 409 (1996).

\bibitem {Bj89}
J.D. Bjorken, Nucl.\ Phys.\ B (Proc.\ Suppl.) {\bf 11}, 325 (1989).

\bibitem {Heff}
For a review, see:
G. Buchalla, A.J. Buras and M.E. Lautenbacher, Rev.\ Mod.\ Phys.\ {\bf
68}, 1125 (1996).

\bibitem {Chen}
H.-Y. Cheng, Phys.\ Lett.\ B {\bf 335}, 428 (1994);
H.-Y.~Cheng and B.~Tseng, Preprint IP-ASTP-04-97 [hep-ph/9708211].

\bibitem {Soar}
J.M. Soares, Phys.\ Rev.\ D {\bf 51}, 3518 (1995).

\bibitem {Stech97}
M. Neubert and B. Stech, Preprint CERN-TH/97-99 [hep-ph/9705292], to
appear in {\sl Heavy Flavours}, Second Edition, edited by A.J.~Buras
and M.~Lindner (World Scientific, Singapore);
M. Neubert, Preprint CERN-TH/97-169 [hep-ph/9707368], to appear in the
Proceedings of the High-Energy Physics Euroconference on Quantum
Chromodynamics (QCD 97), Montpellier, France, July 1997.

\bibitem {CLEOhad}
T.E. Browder, K. Honscheid and D. Pedrini, Ann.\ Rev.\ Nucl.\ Part.\
Sci.\ {\bf 46}, 395 (1996).

\bibitem {Rodri}
J.L Rodriguez, to appear in the Proceedings of the 2nd International
Conference on $B$ Physics and CP Violation, Honolulu, Hawaii, March
1997.

\bibitem {Miller}
D. Miller, these Proceedings.

\bibitem {Besson}
D. Besson, these Proceedings.

\bibitem {Nari94}
S. Narison, Phys.\ Lett.\ B {\bf 322}, 247 (1994).

\bibitem {Huan94}
T. Huang and C.W. Luo, Phys.\ Rev.\ D {\bf 53}, 5042 (1996).

\bibitem {CLEpiK}
CLEO Coll.\ (R. Godang et al.), Preprint CLNS-97-1522 [hep-ex/9711010].

\bibitem {CLEeta}
CLEO Coll., CLEO~CONF~97-22, contributed paper eps97-333 to this
Conference;
CLEO~CONF~97-13 (1997).

\bibitem {Dighe97}
A.S. Dighe, M. Gronau and J.L. Rosner, Phys.\ Rev.\ Lett.\ {\bf 79},
4333 (1997).

\bibitem {CPpiK}
For a review, see: R. Fleischer, Int.\ J.~Mod.\ Phys.~A {\bf 12}, 2459
(1997).

\bibitem {AliGreub}
A. Ali and C. Greub, Preprint DESY-97-126 [hep-ph/9707251];
A.~Ali et al., Preprint DESY-97-235 [hep-ph/9712372].
%A.~Ali, J.~Chay, C.~Greub and P.~Ko, Preprint DESY-97-235
%[hep-ph/9712372].

\bibitem {Desh97}
N.G. Deshpande, B. Dutta and S. Oh, Preprint OITS-641 [hep-ph/9710354].

\bibitem {BuFl}
A.J. Buras and R. Fleischer, Phys.\ Lett.\ B {\bf 341}, 379 (1995).

\bibitem {charmp}
M. Ciuchini et al., Nucl.\ Phys.\ B {\bf 501}, 271 (1997);
%M. Ciuchini, E. Franco, G. Martinelli and L. Silvestrini, Nucl.\
%%Phys.\
%B {\bf 501}, 271 (1997);
M. Ciuchini et al., Preprint CERN-TH/97-188 [hep-ph/9708222].

\bibitem {BjHawaii}
J.D. Bjorken, Preprint SLAC-PUB-7521 [hep-ph/9706524], to appear in the
Proceedings of the 2nd International Conference on $B$ Physics and CP
Violation, Honolulu, Hawaii, March 1997.

\bibitem {resc1}
A.J. Buras, R. Fleischer and T. Mannel, Preprint CERN-TH/97-307
[hep-ph/9711262].

\bibitem {resc2}
M. Neubert, Preprint CERN-TH/97-342 [hep-ph/9712224].

\bibitem {resc3}
A.F. Falk et al., Preprint JHU-TIPAC-97018 [hep-ph/9712225].
%A.F. Falk, A.L. Kagan, Y. Nir and A.A. Petrov, Preprint
%%JHU-TIPAC-97018
%[hep-ph/9712225].

\bibitem {resc4}
D. Atwood and A. Soni, Preprint [hep-ph/9712287];
these Proceedings.

\bibitem {AtSoni}
D. Atwood and A. Soni, Phys.\ Lett.\ B {\bf 405}, 150 (1997).

\bibitem {HouTs}
W.-S. Hou and B. Tseng, Preprint HEPPH-9705304 [hep-ph/9705304];
X.-G. He, W.-S. Hou and C.S. Huang, Preprint [hep-ph/9712478].

\bibitem {Frit}
H. Fritzsch, Preprint CERN-TH-97-200 [hep-ph/9708348].

\bibitem {KaPetr}
A.L. Kagan and A.A. Petrov, Preprint UCHEP-27 [hep-ph/9707354].

\bibitem {Halp}
I. Halperin and A. Zhitnitsky, Phys.\ Rev.\ D {\bf 56}, 7247 (1997);
Preprint HEPPH-9705251 [hep-ph/9705251].

\bibitem {Shur97}
E.V. Shuryak and A.R. Zhitnitsky, Preprint NI-97033-NQF
[hep-ph/9706316].

\bibitem {Yuan}
F. Yuan and K.-T. Chao, Phys.\ Rev.\ D {\bf 56}, 2495 (1997).

\bibitem {Petrov}
A.A. Petrov, Preprint JHU-TIPAC-97016 [hep-ph/9712497].

\bibitem {Ahmady}
M.R. Ahmady, E. Kou and A. Sugamoto, Preprint RIKEN-AF-NP-274
[hep-ph/9710509].

\bibitem {DuKim97}
D. Du, C.S. Kim and Y. Yang, Preprint BIHEP-TH-97-15 [hep-ph/9711428];
D.~Du and M.~Yang, Preprint BIHEP-TH-97-17 [hep-ph/9711272].

\bibitem {CLEMXs}
CLEO Coll., CLEO~CONF~97-13, contributed paper to this Conference.

\bibitem {Lipkin}
H.J. Lipkin, Preprint ANL-HEP-CP-97-45A [hep-ph/9708253].

\bibitem {Cheng97}
H.-Y. Cheng and B. Tseng, Preprint IP-ASTP-03-97 [hep-ph/9707316].

\bibitem {Datta}
A. Datta, X.G. He and S. Pakvasa, Preprint UH-511-864-97
[hep-ph/9707259].

\end{thebibliography}
\end{document}